\documentclass[11pt,a4paper]{article}
\usepackage{jheppub_kim}

\usepackage{pdflscape}
\usepackage{amsmath}
\usepackage{amssymb}
\usepackage{dcolumn}
\usepackage{bm}
\usepackage{color}
\usepackage{epsfig}
\usepackage{amsfonts}
\usepackage{graphicx}
\usepackage{subfigure}
\usepackage{dcolumn}

\newcommand{\be}{\begin{equation}}
\newcommand{\ee}{\end{equation}}
\newcommand{\bea}{\begin{eqnarray}}
\newcommand{\eea}{\end{eqnarray}}

\def\be{\begin{equation}}
\def\ee{\end{equation}}
\def\bea{\begin{eqnarray}}
\def\eea{\end{eqnarray}}

\begin{document}

\title{ Black String in Massive Gravity}

\author[a]{Seyed Hossein Hendi,}
\author[b]{Hayedeh Zarei,}
\author[c]{Mir Faizal,}
\author[d]{Behnam Pourhassan,}
\author[e]{Zahra Armanfard}

\affiliation[a] {Physics
Department and Biruni Observatory, College of Sciences, Shiraz
University, Shiraz 71454, Iran.\\
Canadian Quantum Research Center 204-3002 32 Ave Vernon, BC V1T
2L7 Canada.} \affiliation[b] {Physics Department and Biruni
Observatory, College of Sciences, Shiraz University, Shiraz 71454,
Iran.} \affiliation[c] {Department of Physics and Astronomy,
University of Lethbridge, Lethbridge, Alberta, T1K 3M4, Canada.\\
Irving K. Barber School of Arts and Sciences,
University of British Columbia, Kelowna, British Columbia, V1V
1V7, Canada.\\
Canadian Quantum Research Center, 204-3002 32 Ave Vernon, BC V1T 2L7 Canada.}
\affiliation[d] {Iran Science Elites Federation, Tehran, Iran.\\
Canadian Quantum Research Center 204-3002 32 Ave Vernon, BC V1T
2L7 Canada.} \affiliation[e] {Department of physics and astronomy,
Washington State University, Pullman, Washington 99164-2814, USA}

\emailAdd{hendi@shirazu.ac.ir (corresponding author)}
\emailAdd{mirfaizalmir@googlemail.com}
\emailAdd{b.pourhassan@candqrc.ca (corresponding author)}

\abstract{We will analyze a black string in dRGT massive theory of
gravity. Considering different approaches, we will study the
critical behavior, phase transition and thermal stability for such
a black string solution. We will also analyze the Van der Waals
behavior for this system, and observe that the Van der Waals
behavior depends on the graviton mass.  It will  be observed that
the thermal fluctuations can modify the behavior of this system.
We will explicitly analyze the effect of such fluctuations  on the
stability of this  black string solution. This will be done using
the  Hessian matrix for this system. }


\maketitle

\section{Introduction}
Even though the general relativity is one of the most well tested
theories \cite{Will2005}, it is possible to be corrected in   both
the IR   \cite{Isham}) and the UV
\cite{Weinberg89,Martin12,Ishak19}) limits. This has motivated the
construction of a massive theory of gravity, in which the
gravitons have a  mass  in IR limit
\cite{deRham10,deRham11,deRham14, deRham11,Hassan12}. The results
from the  LIGO collaboration have constraint the  graviton mass to
$m_g<1.2 \times 10$ $ev/c^2$ \cite{deRham14,LIGO}. However, below
this limit, it is possible for the gravitons to be massive. It may
be noted if gravitons are massive, then it  is possible to obtain
a cosmological constant, which can explain the accelerated
expansion of the universe \cite{Leon,Akrami13,Akrami15}. So, there
is a strong motivation to study such a massive theory of gravity.

Even though it is important to study a theory of massive
gravitons, the original Fierz-Pauli theory of massive gravity
\cite{Fierz1939} is not well-defined for the vanishing limit of
graviton mass, due to the vDVZ discontinuity
\cite{Dam1970,Zakharov70,Nieuwenhuizen1973}. However, using the
Vainshtein mechanism it is possible to  obtain  a non-linear
theory of  massive gravity (Stueckelberg trick)
\cite{Vainshtein,Hinterbichler}. Even though this non-linear
theory of massive gravity does not have the vDVZ discontinuity, it
has  ghosts and so  is not a physical theory    \cite{BD}.  It is
possible to obtain a ghost free  massive theory of gravity,
without the  vDVZ discontinuity, and this theory is called as the
dRGT  gravity \cite{deRham10}. Several black hole solutions have
been constructed  using the  dRGT gravity, and it has been
observed that the graviton mass can produce important
modifications to such solutions
\cite{Koyama11,Nieuwenhuizen11,Gruzinov11,Berezhiani11,deRhamPLB12,Hendi2016,Suchant16,Do93,Do94,Hendi2017}.
It has also been observed that   the graviton mass can have
important consequences   for the thermodynamics of various  black
hole solutions
\cite{HendiPRDMann,DehghaniEPJC,DehghaniCQG,DehghaniPRD}.

It is possible to  study of thermodynamics of  AdS black holes in
an  extended phase space, and this is done by identifying the
cosmological constant with the   thermodynamic pressure
\cite{exten1, exten2}. It is also possible to study a
 Van der Waals like behavior for such
asymptotically AdS black holes \cite{Dolan11,Kubiznak12}. In fact,
triplet point \cite{Altamirano14,Wei14,Hennigar15} and reentrant
phase transitions \cite{Gunasekaran12,Frassino14,HennigarEntropy}
for such AdS black hole solution have been studied using such a
Van der Waals behavior. It has been also been observed that  the
graviton mass in massive gravity can produce   new interesting
phase transitions in black hole solutions   \cite{HendiPRDMann,
massive2, massive3, massive4}. So, it is important to study Van
der Waals behavior for black hole solutions in massive gravity.

It is possible to extended the black hole solution to a black
string solution \cite{Lemos}. It is also possible to consider
black string solutions in asymptotically AdS space-time, and hence
the CFT dual to such solutions can also be studied   \cite{adsst1,
adsst2}.  It may be noted that the Hawking radiation and  greybody
factor of black strings in massive gravity has also  been studied,
and it was observed that this greybody factor  depends on the
graviton mass in this theory of massive gravity \cite{grey}. In
fact, a solution for rotating black strings has also been
constructed in massive gravity, and the dependence of the
thermodynamics on the mass of graviton has also been studied, for
such a solution \cite{main}. It may be noted that black strings
have an interesting behavior, and it has been observed that they
can even produce effects, which are not observed in ordinary black
holes
\cite{Horne92,Gregory93,Emparan2000,Gregory2000,Wiseman03,Sorkin04,Kudoh05,Sorkin06,Kleihaus,Bogdanos09,Hendi10,Figueras,Hendi13,Kalisch,Emparan18,Cisterna18,Kanti,Cisterna19,Nakas}.
As it is possible to study the Van der Waals behavior for AdS
black hole in massive gravity \cite{HendiPRDMann,  massive3}, we
will analyze the Van der Waals like behavior  for a black string
in massive gravity.

It may be noted that the thermodynamics of black holes can get
corrected due to thermal fluctuations \cite{cjp0, CJP}. It has
also been observed that these thermal fluctuations are in the
thermodynamics of the black holes are produced by the quantum
fluctuation in the geometry of a black hole \cite{cjp1, cjp2,
cjp4, EPL, NPB}. It has been possible to study the effects of such
thermal fluctuations on the  Van der Waals like behavior of black
holes \cite{ther1, ther2}.  Thus, it is  possible that these
thermal fluctuations can change the behavior of various black hole
solutions.  As black strings are important solutions, we will
analyze the effects of thermal fluctuations on black strings in
massive theory of gravity.  The effects of thermal fluctuations on
the thermodynamics  of black holes has been studied in massive
gravity \cite{massive2, massive4}. Furthermore,   the stability of
black hole solutions, and the effects of thermal fluctuations on
the stability of  black holes can be studied using the Hessian
matrix \cite{Hess1, Hess2, Hess3}.  So, we will analyze the
effects of thermal fluctuations on the stability of a  black
string in massive gravity using this Hessian matrix.

\section{ Black String Solution  \label{model}}

Now we give a brief review of the field equation and rotating
solutions of dRGT nonlinear massive gravity, which is free of the
ghost. The modification of the Einstein-Maxwell field equations
from  the graviton mass  can be written as
\cite{deRham11,Vegh}
\begin{equation}
G_{\mu \nu }+m_{g}^{2}\mathcal{X}_{\mu \nu }=-\frac{1}{2}{g_{\mu \nu
}F^{\alpha \beta }}{F_{\alpha \beta }}+2{F_{\mu \lambda }}{F_{\nu }}%
^{\,\lambda },  \label{Feq1}
\end{equation}%
\begin{equation}
{\nabla _{\mu }}{F^{\mu \nu }}=0,  \label{Feq2}
\end{equation}%
where $m_{g}$ is the graviton mass parameter, $F_{\mu \nu }$ denotes the
Faraday tensor which is constructed using the gauge field ${A_{\nu }}$ as ${%
F_{\mu \nu }}={\partial _{\lbrack \mu }}{A_{\nu ]}}$ and
\begin{eqnarray}
\mathcal{X}_{\mu \nu } &=&\mathcal{K}_{\mu \nu }-\alpha \left( \mathcal{K}%
_{\mu \nu }^{2}-\mathcal{K}\mathcal{K}_{\mu \nu }\right) +3\beta \left(
\mathcal{K}_{\mu \nu }^{3}-\mathcal{K}\mathcal{K}_{\mu \nu }^{2}+\frac{%
\mathcal{K}^{2}-[\mathcal{K}^{2}]}{2}\mathcal{K}_{\mu \nu }\right)  \notag \\
&&-g_{\mu \nu }\left( \mathcal{K}+\frac{\alpha \left( \mathcal{K}^{2}-[%
\mathcal{K}^{2}]\right) +\beta \left( \mathcal{K}^{3}-3\mathcal{K}[\mathcal{K%
}^{2}]+2[\mathcal{K}^{3}]\right) }{2}\right).  \label{X}
\end{eqnarray}%
Here $\mathcal{K}_{\nu }^{\mu }=\delta _{\nu }^{\mu }-\left( \sqrt{g^{-1}%
\tilde{f}}\right) _{\nu }^{\mu }$ and $[\mathcal{K}^{n}]=(\mathcal{K}%
^{n})_{\alpha }^{\alpha }$. Now $\tilde{f}$ is the trace of
the metric $\tilde{f}_{\mu \nu }=f_{ab}\partial _{\mu }\phi ^{a}\partial
_{\nu }\phi ^{b}$, with  $f_{ab}$ as the reference (fiducial) metric and $%
\phi ^{a}$'s as the St\"{u}ckelberg scalars (which are introduced
to restore general covariance of the theory). The choices of the
reference metric can be used to obtain  different subclasses of
dRGT massive gravity. In this paper, the unitary gauge $\phi
^{a}=x^{\mu }\delta _{\mu }^{a}$ \cite{Vegh} is chosen so that the
observable (physical) metric tensor $g_{\mu \nu }$ can describe
the five degrees of freedom of the massive graviton.

Now, we begin with the following metric of rotating black string
space-time in dRGT massive gravity
\begin{equation}
ds^{2}=\left[ \frac{r^{2}\omega _{0}^{2}}{l^{2}}-f(r)\right] dt^{2}+\frac{%
dr^{2}}{f(r)}+2l\omega _{0}\left[ f(r)-\frac{r^{2}}{l^{2}}\right]
dtd\varphi +r^{2}\left[ 1-\frac{l^{2}\omega
_{0}^{2}}{r^{2}}f(r)\right] d\varphi
^{2}+r^{2}d\boldsymbol{z}^{2}, \label{Metric}
\end{equation}%
where  $\boldsymbol{z}=z/l$ is the dimensionless coordinates along
the black string and  $\omega _{0}$ is  rotation parameter (one
can obtain the static line element with $\omega _{0}=0$). We will
work in the cylindrical coordinates, with $-\infty <t<+\infty $, $0\leq r<+\infty $, $-\infty <z<+\infty $ and $0\leq \varphi <2\pi $. The consistent gauge potential is defined as $A_{\mu }=\left[ h(r),0,-h(r)l%
\omega _{0},0\right] $, where $h(r)$ is an arbitrary function of the radial
coordinate $r$, and its explicit functional form can be obtained from the
Maxwell equations (\ref{Feq2}). The exact solutions of the field equations
with the mentioned metric (\ref{Metric}) are given by \cite{main}
\begin{eqnarray}
f(r) &=&-\frac{4m}{r}+\frac{4q^{2}}{r^{2}}-\frac{\Lambda r^{2}}{3}%
-c_{1}r+c_{0},  \label{f(r)} \\
h(r) &=&-\frac{2q}{r},  \label{h(r)}
\end{eqnarray}%
where two integration constants $m$ and $q$ are related to the ADM
mass and the electric charge per unit $\boldsymbol{z}$, along $z$
direction.  We can define  the following     constants in this
theory of massive gravity
\begin{equation}
3m_{g}^{2}\left( 1+\alpha +\beta \right) \equiv -\Lambda ,\,\
\,\,m_{g}^{2}b_{0}(1+2\alpha +3\beta )\equiv c_{1},\text{ \ \ }%
m_{g}^{2}b_{0}^{2}(\alpha +3\beta )\equiv c_{0}.  \label{Constants}
\end{equation}%
To  obtain this exact solution of the
field equations, the    reference metric can be written as
\begin{equation}
f_{\mu \nu }=\left(
\begin{array}{cccc}
\frac{b_{0}^{2}\omega _{0}^{2}}{l^{2}} & 0 & -\frac{b_{0}^{2}\omega _{0}}{l}
& 0 \\
0 & 0 & 0 & 0 \\
-\frac{b_{0}^{2}\omega _{0}}{l} & 0 & b_{0}^{2} & 0 \\
0 & 0 & 0 & \frac{b_{0}^{2}}{l^{2}}%
\end{array}%
\right).  \label{fiducial}
\end{equation}%
In order to investigate the geometric properties of the solutions, we can
calculate the curvature scalars. Considering the metric (\ref{Metric}), we
can obtain%
\begin{eqnarray*}
R_{\alpha \beta \gamma \delta }R^{\alpha \beta \gamma \delta } &=&\left(
\frac{d^{2}f(r)}{dr^{2}}\right) ^{2}+\left( \frac{2}{r}\frac{df(r)}{dr}%
\right) ^{2}+\left( \frac{2f(r)}{r^{2}}\right) ^{2}, \\
R_{\alpha \beta }R^{\alpha \beta } &=&\frac{1}{2}\left( \frac{d^{2}f(r)}{%
dr^{2}}+\frac{2}{r}\frac{df(r)}{dr}\right) ^{2}+2\left( \frac{1}{r}\frac{%
df(r)}{dr}+\frac{f(r)}{r^{2}}\right) ^{2}, \\
R &=&\frac{d^{2}f(r)}{dr^{2}}+\frac{4}{r}\frac{df(r)}{dr}+\frac{2f(r)}{r^{2}}.
\end{eqnarray*}

\begin{figure}[h!]
 \begin{center}$
 \begin{array}{cccc}
\includegraphics[width=60 mm]{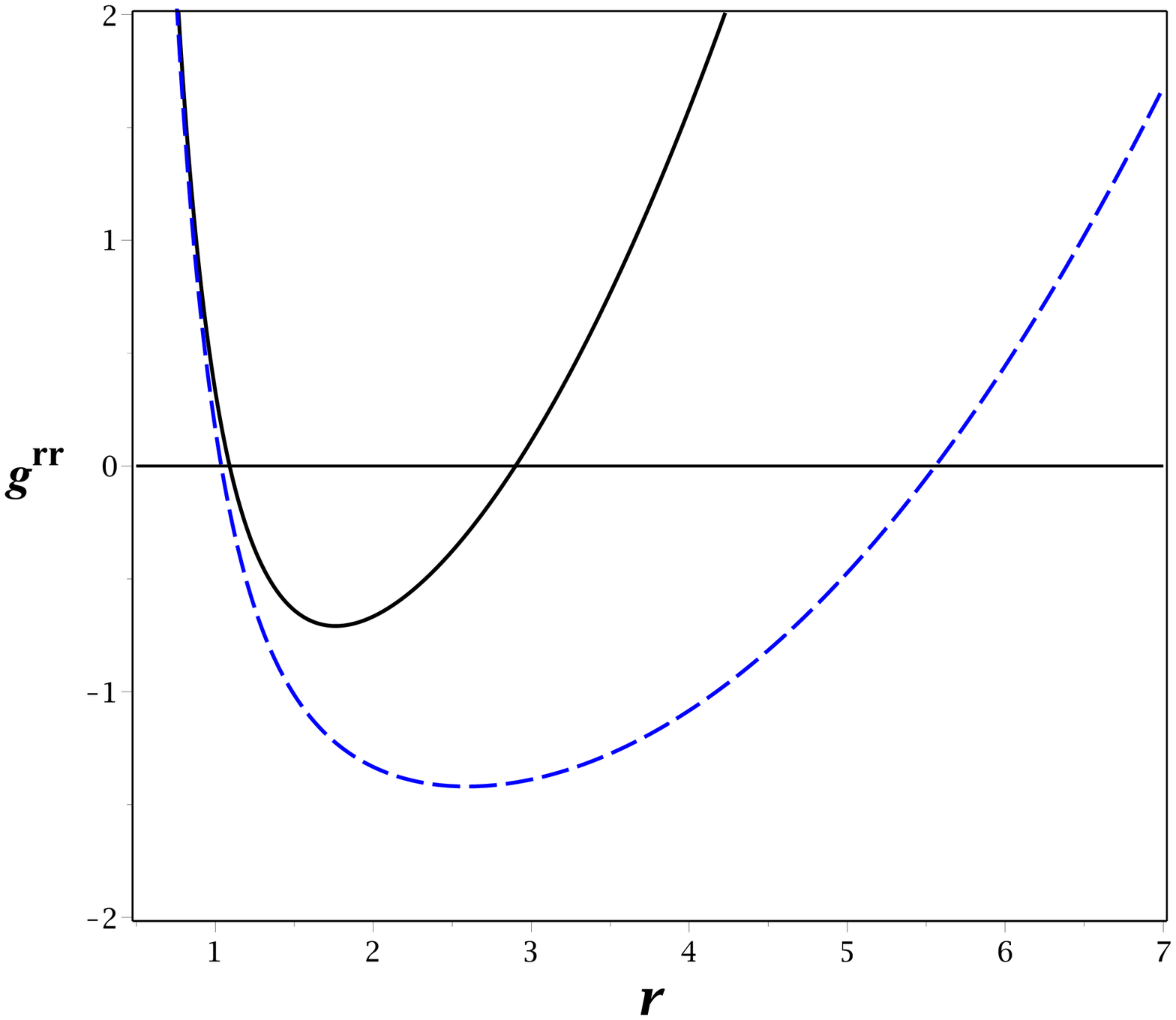}\includegraphics[width=60 mm]{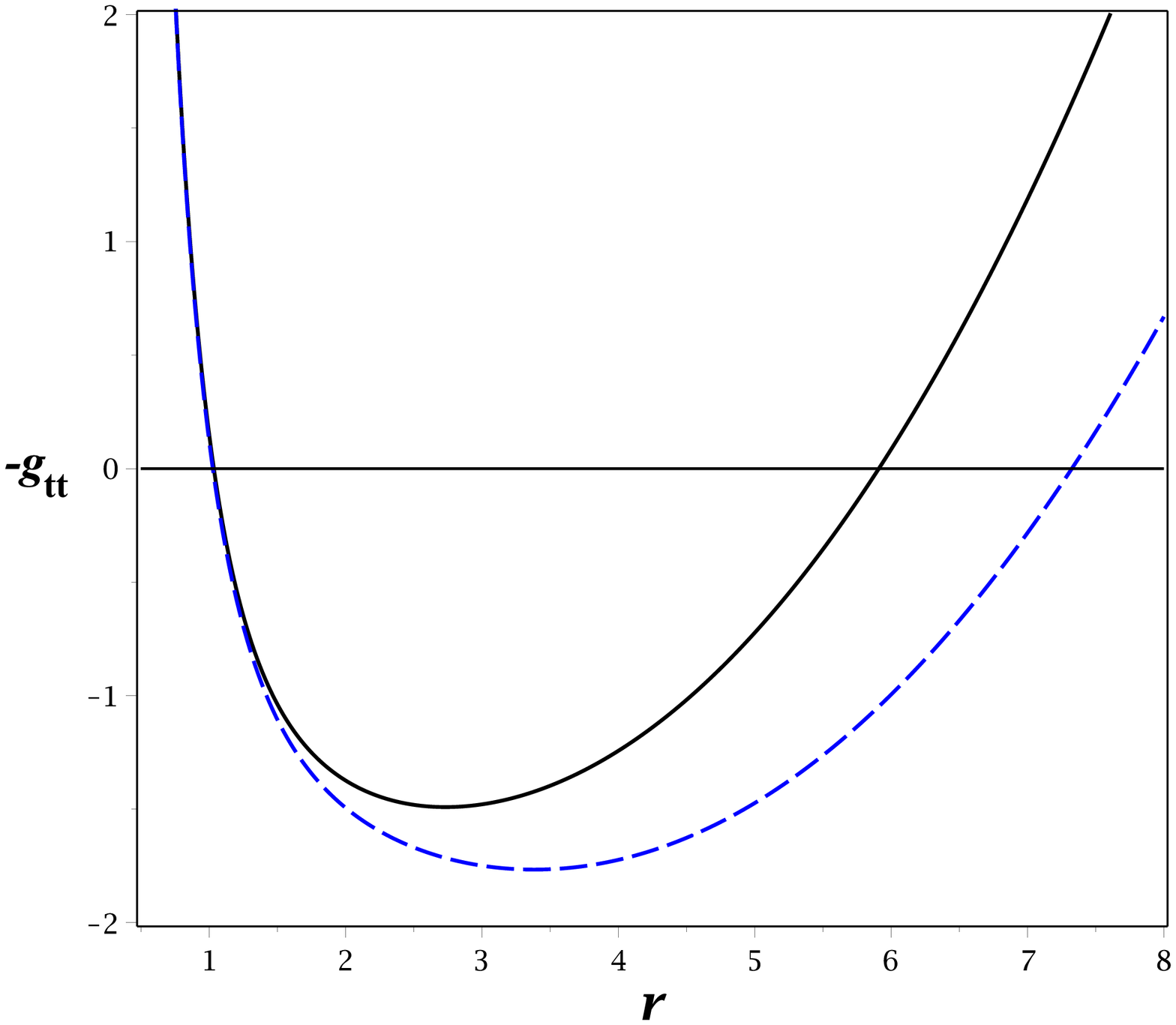}
 \end{array}$
 \end{center}
\caption{$g^{rr}$ (left) and $-g_{tt}$ (right) versus $r$ for
$c=c_0=c_1=q=m=1$. \textbf{Left panel: } $\Lambda=-1$ (continuous
line) and $\Lambda=-0.5$ (dashed line). \textbf{Right panel:}
$\Lambda=-0.5$ and $l=1$ with $\protect\omega_{0}=0.1$ (continuous
line) and $\protect\omega_{0} =0.2$ (dashed line).}
 \label{grr-gtt}
\end{figure}

After inserting $f(r)$ into the above relations, we can find that
near the origin ($r\rightarrow 0^{+}$)
\begin{eqnarray*}
R_{\alpha \beta \gamma \delta }R^{\alpha \beta \gamma \delta } &\approx
&r^{-8}, \\
R_{\alpha \beta }R^{\alpha \beta } &\approx &r^{-8}, \\
R &\approx &r^{-2},
\end{eqnarray*}%
while far from the black object ($r\rightarrow \infty $), we have
\begin{eqnarray*}
R_{\alpha \beta \gamma \delta }R^{\alpha \beta \gamma \delta } &\approx &%
\frac{8\Lambda ^{2}}{3}, \\
R_{\alpha \beta }R^{\alpha \beta } &\approx &4\Lambda ^{2}, \\
R &\approx &4\Lambda.
\end{eqnarray*}%
Thus, it seems that there is a curvature singularity along $z-$direction ($r=0
$). Due to the Cosmic Censorship Conjecture,  there should be  an  event horizon
for such a singularity, and   we can  interpret such an event horizon  as a black string. Using  $g^{rr}=0
$, we can find the location of the event horizon as
\begin{equation}
g^{rr}=-\frac{\Lambda }{3}r_{+}^{2}-c_{1}r_{+}+c_{0}-\frac{4m}{r_{+}}+\frac{%
4q^{2}}{r_{+}^{2}}=0.  \label{Horizon}
\end{equation}%
Here  the event horizon is the largest real positive root of Eq. (\ref%
{Horizon}), with positive slope. Since we cannot obtain an analytical form for
$r_{+}=r_{+}(c_{0},c_{1},m,q,\Lambda )$, we plot Fig. \ref{grr-gtt} (left
panel) to confirm the existence of such an event horizon. According to this
figure, one finds the curvature singularity may be covered by an event
horizon and the radius of the event horizon depends on the free parameters in the theory.

Another interesting properties  of black holes is  the infinite
redshift surface. Although for static black holes the surface of
event horizon coincides the static limit surface (infinite
redshift), the situation is different for the rotating solutions.
In order to find the static limit surface with radius $r_{i}$, one
has to consider $g_{tt}=0$
\begin{equation}
g_{tt}=\left( \frac{\Lambda }{3}+\frac{\omega_{0}
^{2}}{l^2}\right)
r_{i}^{2}+c_{1}r_{i}-c_{0}+\frac{4m}{r_{i}}-\frac{4q^{2}}{r_{i}^{2}}=0
\label{SLS}
\end{equation}%
which is  different from Eq. (\ref{Horizon}). Considering the right
panel of Fig. \ref{grr-gtt}, it is clear that the radius of static limit
surface (the largest real positive root of Eq. (\ref{SLS})) is affected by
the rotation parameter.

Taking into account the dominant term of the metric function for
asymptotically large $r$ ($\frac{\Lambda r^{2}}{3}-$term), one may
deduce that the rotating black string approach the AdS space-time.
However, we should note that although the behavior of the
curvature invariant support  such a statement, the asymptotic
symmetry group is not necessarily that of pure AdS (see
\cite{Cai2015} for a counterexample).

\section{Thermodynamics  \label{Thermo}}

Here, we are going to calculate thermodynamic and conserved
quantity for the charged rotating black string. It is obvious that
the cylindrically symmetric metric (\ref{Metric}) is extended
infinitely along $z-$direction. So,  both the mass and charge are
distributed along $z-$direction, and would be infinite. So, to
obtain the physical finite quantities, we can  calculate the mass
and charge linear densities, which are the mass and charge per
unit length along $z-$direction.

Applying the Brown-York approach \cite{Brown:1992br}, and
rewriting the metric in the canonical form, one can calculate the
conserved quantities for this system. So, using  the two known
Killing vectors $\partial /\partial t$ and $\partial /\partial
\varphi $ of the rotating metric (\ref{Metric}), we can calculate
their corresponding conserved charges. These conserved charges are
mass and angular momentum (per unit $\boldsymbol{z}$ along
$z-$direction). They are   associated with the  time translation
and rotation invariance of the system.
\begin{eqnarray}
M &=&\left\vert 1-\omega _{0}^{2}\right\vert m,  \label{Mass} \\
J &=&\frac{\left\vert 1-\omega _{0}^{2}\right\vert m}{l}\omega
_{0},  \label{AngMom}
\end{eqnarray}%
where these relations reduce to $M=m$ and $J=0$ for the static case ($\omega
_{0}=0$). Using the  axial symmetric  of the rotating metric (\ref%
{Metric}), one can obtain  $\xi ^{\mu }=\left( 1,0,\Omega _{H},0\right) $\ as
the Killing vector, for which the angular velocity at the horizon is
\begin{equation}
\Omega _{H}=-\left. \frac{g_{t\varphi }}{g_{\varphi \varphi }}\right\vert
_{r=r_{+}}=\frac{\omega _{0}}{l}.  \label{AngVel-a}
\end{equation}

Now, we can  calculate the electric charge and the  electric
potential for this system.  The electric charge density (per unit
$\boldsymbol{z}$ along $z-$direction) of black string can be
calculated using the Gauss law as
\begin{equation}
Q=q.   \label{Charge}
\end{equation}%
The electric potential measured at infinity, with respect to the
horizon, is given by
\begin{equation}
\Phi =\left. A_{\mu }\xi ^{\mu }\right\vert _{r\longrightarrow \infty
}-\left. A_{\mu }\xi ^{\mu }\right\vert _{r\longrightarrow r_{+}}=\frac{%
2\left\vert 1-\omega _{0}^{2}\right\vert q}{r_{+}}.  \label{Pot}
\end{equation}%
So,  the entropy of a black object equals one-quarter of its horizons
area. As this  is a  universal relation between area and entropy, it  holds for all black objects, including
black strings
\cite{AreaLaw,AreaLaw1,AreaLaw2,AreaLaw3,AreaLaw4,AreaLaw5}. The entropy of the
black string per unit $\boldsymbol{z}$ along $z-$direction can be written as%
\begin{equation}
S=\frac{A}{4}=\frac{\pi r_{+}^{2}}{2}  \label{Entropy}
\end{equation}%
The Hawking temperature of black holes can be calculated from its
surface gravity.  Using the vanishing of  the metric function at
the event horizon, we obtain
\begin{eqnarray}
T &=&\frac{1}{2\pi }\sqrt{-\frac{1}{2}\left( \nabla _{\mu }\xi _{\nu
}\right) \left( \nabla ^{\mu }\xi ^{\nu }\right) }=\frac{\left\vert 1-\omega
_{0}^{2}\right\vert }{4\pi }\left. \frac{df(r)}{dr}\right\vert _{r=r_{+}}
\notag \\
&=&\frac{\left\vert 1-\omega _{0}^{2}\right\vert }{4\pi r_{+}}\left(
-\Lambda r_{+}^{2}-2c_{1}r_{+}+c_{0}-\frac{4q^{2}}{r_{+}^{2}}\right) .
\label{Temp}
\end{eqnarray}
Using  these conserved quantities, it is
straightforward to write  the first law of thermodynamics as  \cite{main},
\begin{equation}
dM=TdS+\Omega _{H}dJ+\Phi dQ.  \label{First-a}
\end{equation}

\subsection{Phase Transition}

In the usual form  for the first law of black
hole thermodynamics, the work term  ($P-V$) is absent. However, it is possible to consider such a work term in extended phase space
\cite{Fang}. This is done by identifying the negative cosmological constant with the  thermodynamic pressure
\begin{equation}
P=\frac{-\Lambda }{8\pi }\left\vert 1-\omega _{0}^{2}\right\vert .  \label{P}
\end{equation}
Using the  Eq. (\ref{Temp}), with such a  pressure, we
can obtain  the following equation of state
\begin{equation}
P=\frac{1}{8\pi r_{+}^{2}}\left[ \left( 2c_{1}r_{+}-c_{0}+\frac{4q^{2}}{%
r_{+}^{2}}\right) \left\vert 1-\omega _{0}^{2}\right\vert +4\pi r_{+}T\right].  \label{Pressure}
\end{equation}%
Here the  conjugate variable to pressure  is interpreted as the thermodynamic volume
\begin{equation}
V=\frac{2\pi r_{+}^{3}}{3}.  \label{Volume}
\end{equation}%
This  is equivalent to $\int {4Sdr_{+}}=\frac{2\pi r_{+}^{3}}{%
3}$.

Such an identification leads to an extended version of the first
law of black hole thermodynamics with a volume-pressure term
\begin{equation}
dM=TdS+\Phi dQ+PdV+\mathcal{C}dc_{1},  \label{First-b}
\end{equation}%
where $\mathcal{C}=\left( \frac{\partial M}{\partial c_{1}}\right) _{S,Q,V}=%
\frac{-\left\vert 1-\omega _{0}^{2}\right\vert r_{+}^{2}}{4}$.

\begin{figure}[h!]
 \begin{center}$
 \begin{array}{cccc}
\includegraphics[width=46 mm]{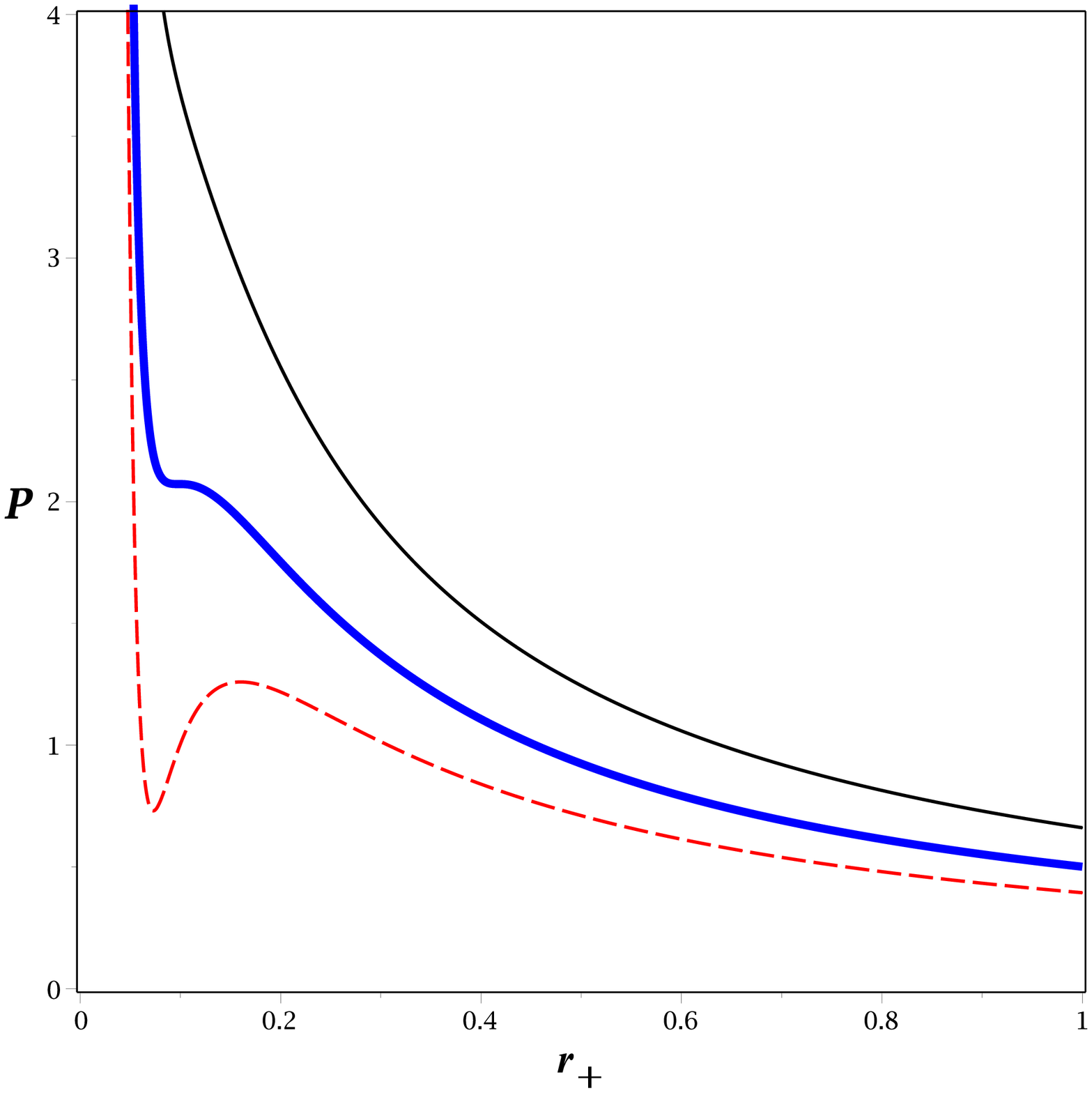}\includegraphics[width=50 mm]{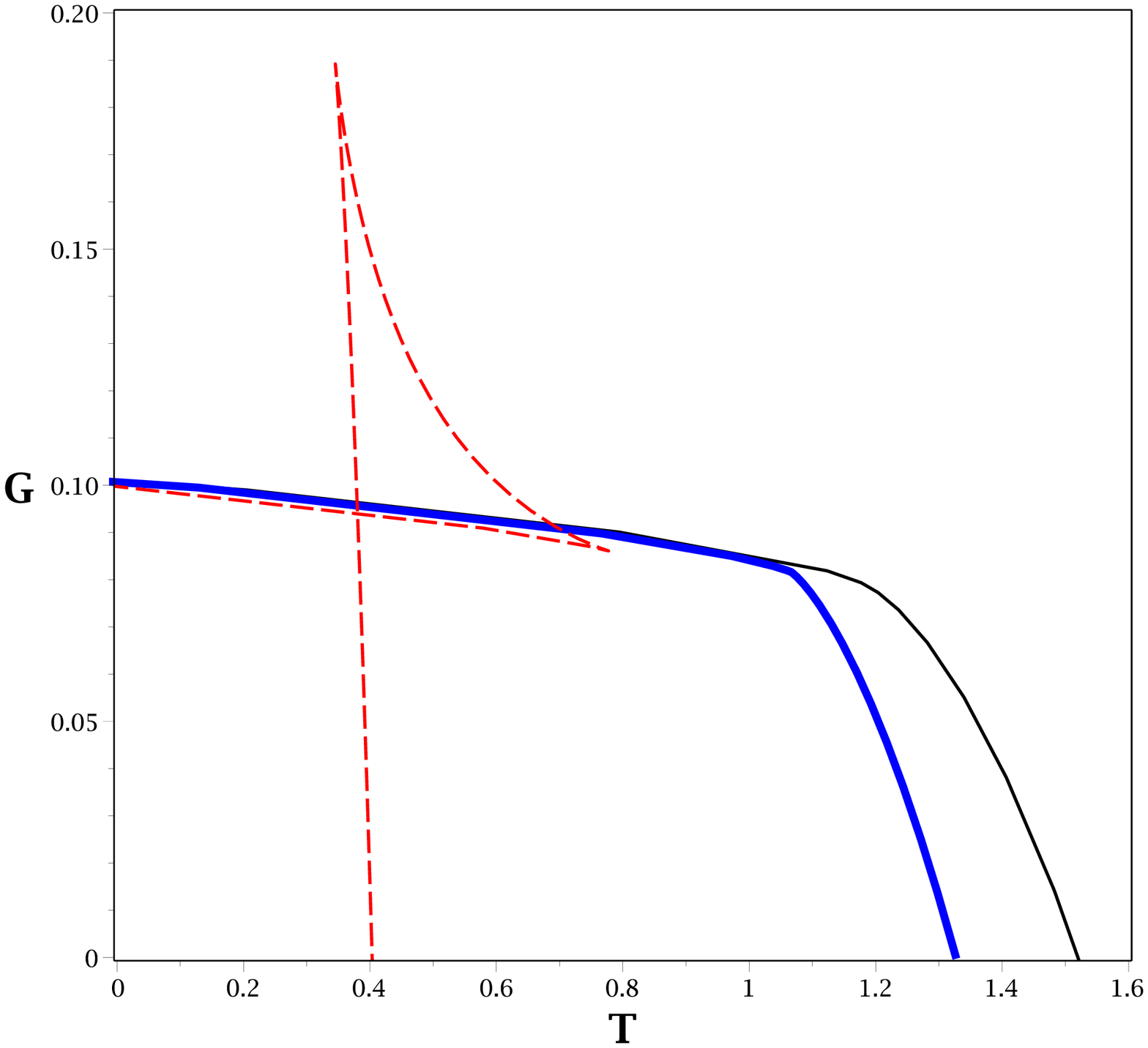}\includegraphics[width=51 mm]{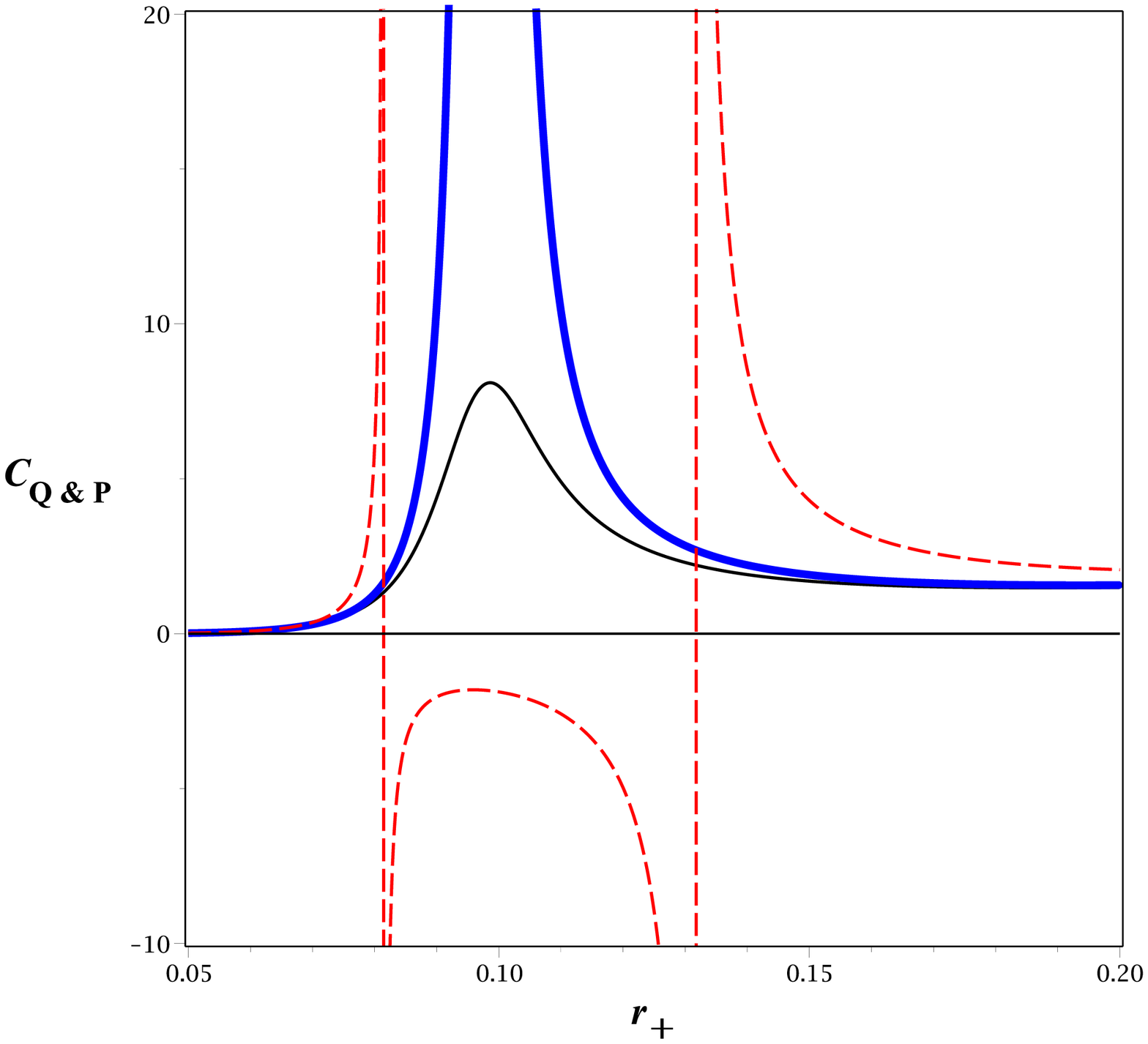}
 \end{array}$
 \end{center}
\caption{$P-r_{+}$ (left), $G-T$ (middle) and $C_{Q,P}-r_{+}$
(right) diagrams for $\protect\omega_{0}=0.2$, $c_1=0.1$, $c_0=1$,
$m=2$, and $q=0.1$.\newline \textbf{Left panel:} $T>T_{c}$
(continuous black line), $T=T_{c}$ (bold blue line) and $T<T_{c}$
(dashed red line).\newline
\textbf{Middle and right panels:} $P>P_{c}$ (continuous black line), $%
P=P_{c} $ (bold blue line) and $P<P_{c}$ (dashed red line).}
 \label{PV-GT-CQ}
\end{figure}

In order to investigate the Van der Waals like behavior of the black string
solution, we can focus on the isothermal $P-V$ diagram (see left panel of
Fig. \ref{PV-GT-CQ}). As  the inflection point of
isothermal curves in $P-V$ diagram determine the critical point, we can use
the following relations%
\begin{equation}
\left( \frac{\partial P}{\partial v}\right) _{T}=0,\text{ \ \ \ \ }\left(
\frac{\partial ^{2}P}{\partial v^{2}}\right) _{T}=0.  \label{dP-ddP}
\end{equation}%
Here  the specific volume $v$  is proportional to $r_{+}$
(here $v=2r_{+}$ in relativistic unit with $l_{P}=1$) \cite{mann}.
Using  the both relations of Eq. (\ref{dP-ddP}), simultaneously, we
can analytically find the following critical quantities,
\begin{eqnarray*}
T_{c} &=&\frac{\sqrt{6}\left\vert 1-\omega _{0}^{2}\right\vert \left( c_{0}^{%
\frac{3}{2}}-3\sqrt{6}qc_{1}\right) }{36\pi q}, \\
r_{c} &=&\frac{2\sqrt{6}q}{\sqrt{c_{0}}}, \\
P_{c} &=&\frac{c_{0}^{2}\left\vert 1-\omega _{0}^{2}\right\vert }{384\pi
q^{2}}.
\end{eqnarray*}

In order to obtain the physical (positive) critical quantities,
one has to satisfy  the following inequality
\begin{equation*}
c_{1}\leq \frac{\sqrt{6}c_{0}^{\frac{3}{2}}}{18q}.
\end{equation*}
It is also notable that unlike $T_{c}$, both critical pressure and
horizon radius are independent of the rotation parameter and
$c_{1}$. Furthermore, using  Eq. (\ref{Pressure}), one can write
the pressure as
\begin{equation}
P=\frac{1}{8\pi r_{+}^{2}}\left[ \left( \frac{4q^{2}}{r_{+}^{2}}%
-c_{0}\right) \left\vert 1-\omega _{0}^{2}\right\vert +4\pi r_{+}T^{\text{eff%
}}\right],
\end{equation}%
where the effective temperature is defined as
\begin{equation*}
T^{\text{eff}}=T+\frac{\left\vert 1-\omega _{0}^{2}\right\vert c_{1}}{2\pi }.
\end{equation*}%
Here  the new effective critical temperature is given by
\begin{equation*}
T_{c}^{\text{eff}}=\frac{\sqrt{6}\left\vert 1-\omega _{0}^{2}\right\vert
c_{0}^{\frac{3}{2}}}{36\pi q}.
\end{equation*}

Now we can use the Gibbs free energy  to study of phase
transitions. Its behavior as a function of pressure and
temperature can be used to analyze the phase transition  in this
system, along with its equilibrium state. Although the Gibbs free
energy has a different functional dependence on its variables in
different phases, the equilibrium state is the one with the lowest
Gibbs free energy for given values of $T$ and $P$. Furthermore,
the differences in the properties of the phases appear as
discontinuities in some derivatives of the Gibbs free energy.
Strictly speaking if the $n^{th}$ order derivatives are
discontinuous, the phase transition is called $n^{th}$ order. The
Gibbs free energy in the extended phase space is obtained as
\begin{equation}
G =M-TS  = \frac{\left( c_{0}r_{+}^{2}+12q^{2}\right) \left\vert
1-\omega _{0}^{2}\right\vert }{8r_{+}}-\frac{\pi }{3}Pr_{+}^{3}.
\label{Gibbs}
\end{equation}%
The qualitative behavior of Gibbs free energy as an implicit
function of temperature is shown in the middle panel of Fig.
\ref{PV-GT-CQ}. Evidently, the swallow-tail characteristic
demonstrates the first order phase transition (as indicated before
in $P-V$ diagram of Van der Waals like behavior).
\subsection{Thermal stability}

Now we  will investigate the local stability of charged rotating
black string solutions using  the canonical ensemble. This can be
done as the   heat capacity can be used to   determine  not only
the number of the phases for  such a system,  but also their
thermal stability. It may be noted that a black hole is thermally
stable (unstable) if its heat capacity is positive (negative). In
addition, the discontinuities (divergences) in the heat capacity
occur at phase transition. Using  the canonical ensemble in the
extended phase space, both the pressure and total charge are kept
constant. So, the  stability can be analyzed using  positivity of
heat capacity at constant pressure and electric charge. The heat
capacity for black strings under these considerations is
given by%
\begin{equation}
C_{Q,P}=\left( T\frac{\partial S}{\partial T}\right) _{Q,P}=\frac{\pi
r_{+}^{2}\left[ 8\pi Pr_{+}^{4}-\left(
2c_{1}r_{+}^{3}-c_{0}r_{+}^{2}+4q^{2}\right) \left\vert 1-\omega
_{0}^{2}\right\vert \right] }{8\pi Pr_{+}^{4}+\left(
12q^{2}-c_{0}r_{+}^{2}\right) \left\vert 1-\omega _{0}^{2}\right\vert }.
\label{CQ}
\end{equation}

Plotting  the behavior of $C_{Q,P}$ in the right panel of Fig. \ref%
{PV-GT-CQ}, one observes  that for $P>P_{c}$, the heat capacity is
positive, with  $T>0$, and so the system is stable.  Now  for
$P=P_{c}$,  $C_{Q,P}$ has only one divergence point. However,  for
$P<P_{c}$, there are two divergence points for $C_{Q,P}$, and the
system undergoes a phase transition.

\subsection{Geometrical Thermodynamics}
Black hole's phase transition can also be studied using
Geometrical Thermodynamics (GT) method. In this approach, one can
use different metrics in terms of thermodynamic quantities to
build a geometrical space. Then, by calculating the Ricci scalar
and its divergence points of those thermodynamic metrics, we can
investigate the phase transition of interested black hole. It is
expected that the curvature singularities of those spacetimes
relate to the heat capacity divergences and root.

During past few years, there have been different attempts to
introduce such a thermodynamic metric and the well known ones are
Weinhold, Ruppeiner, Quevedo and HPEM metrics. Weinhold used the
second derivative of the internal energy with respect to the
entropy and other extensive parameters to build a thermodynamic
metric on the equilibrium space \cite{wmetric1,wmetric2}.
Following his work, Ruppeiner introduced a metric using the minus
second derivative of entropy with respect to the internal energy
and other extensive parameters, which was conformally related to
Weinhold's metric \cite{rmetric1,rmetric2}. The main problem of
these metrics was that neither of them were Legendre invariant,
making them not suitable for describing the phase transition of
black holes, so Quevedo tried to introduce a metric which was
invariant under Legendre transformation \cite{qmetric1,qmetric2}.
But, even though his proposed metric met the Legendre invariance,
it couldn't provide a flawless mechanism to study phase transition
and thermodynamic properties of some certain black holes through
geometrical thermodynamics \cite{hmetric,hmetric2}. In recent
years another metric has been proposed in Ref. \cite{hmetric}
which seemed to solve the problems that other metrics faced. This
new metric is Legendre invariant and also doesn't have the problem
of extra singularities that are not consistent with any of the
phase transition points and roots of heat capacity. In the
following, we aim to employ these thermodynamic metrics to study
phase transition of our interested black hole and compare their
results to see which metric works the best four case of study.

\begin{figure}[h!]
\begin{center}
$%
\begin{array}{cc}
\includegraphics[width=65 mm]{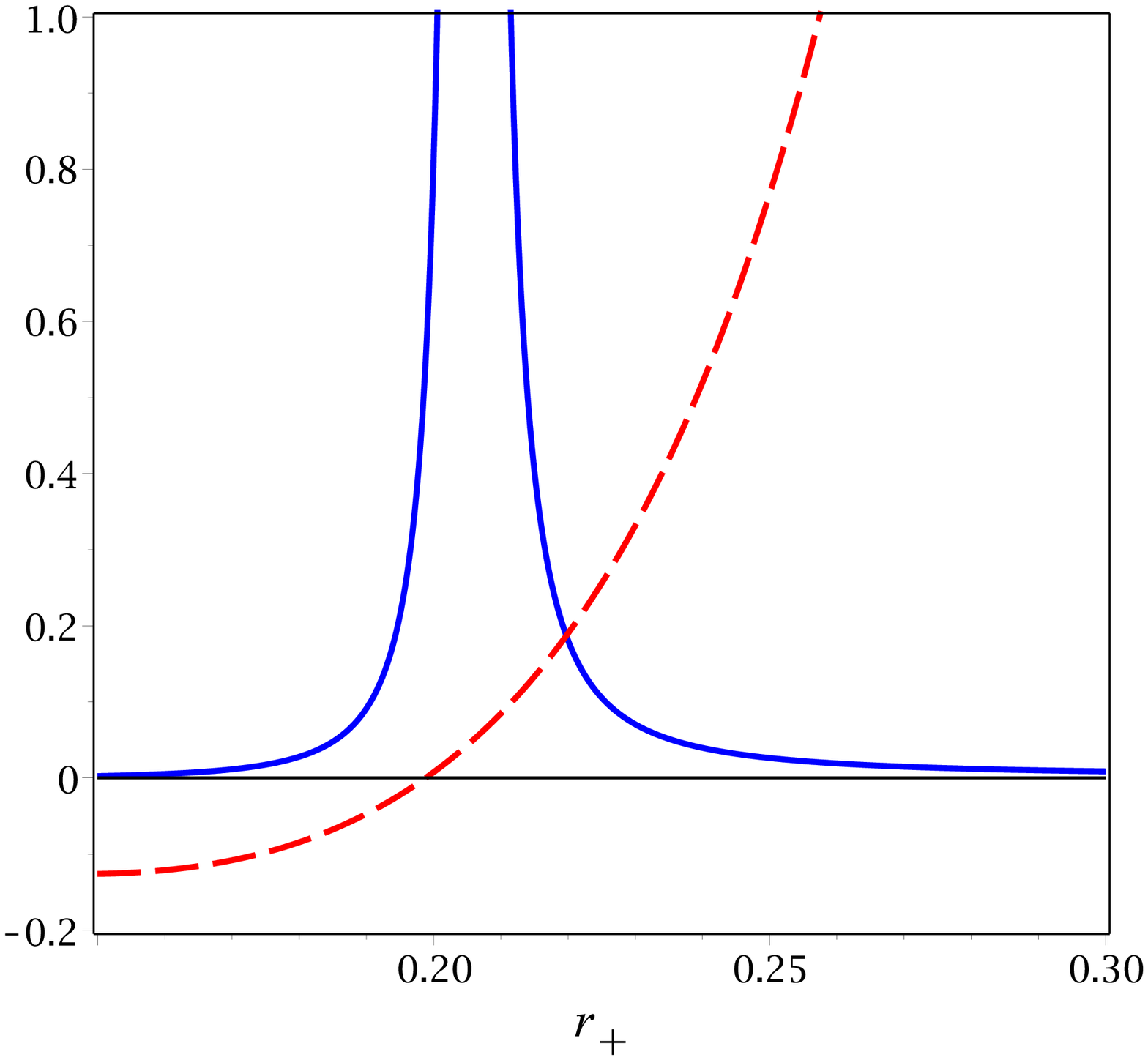} &
\includegraphics[width=65 mm]{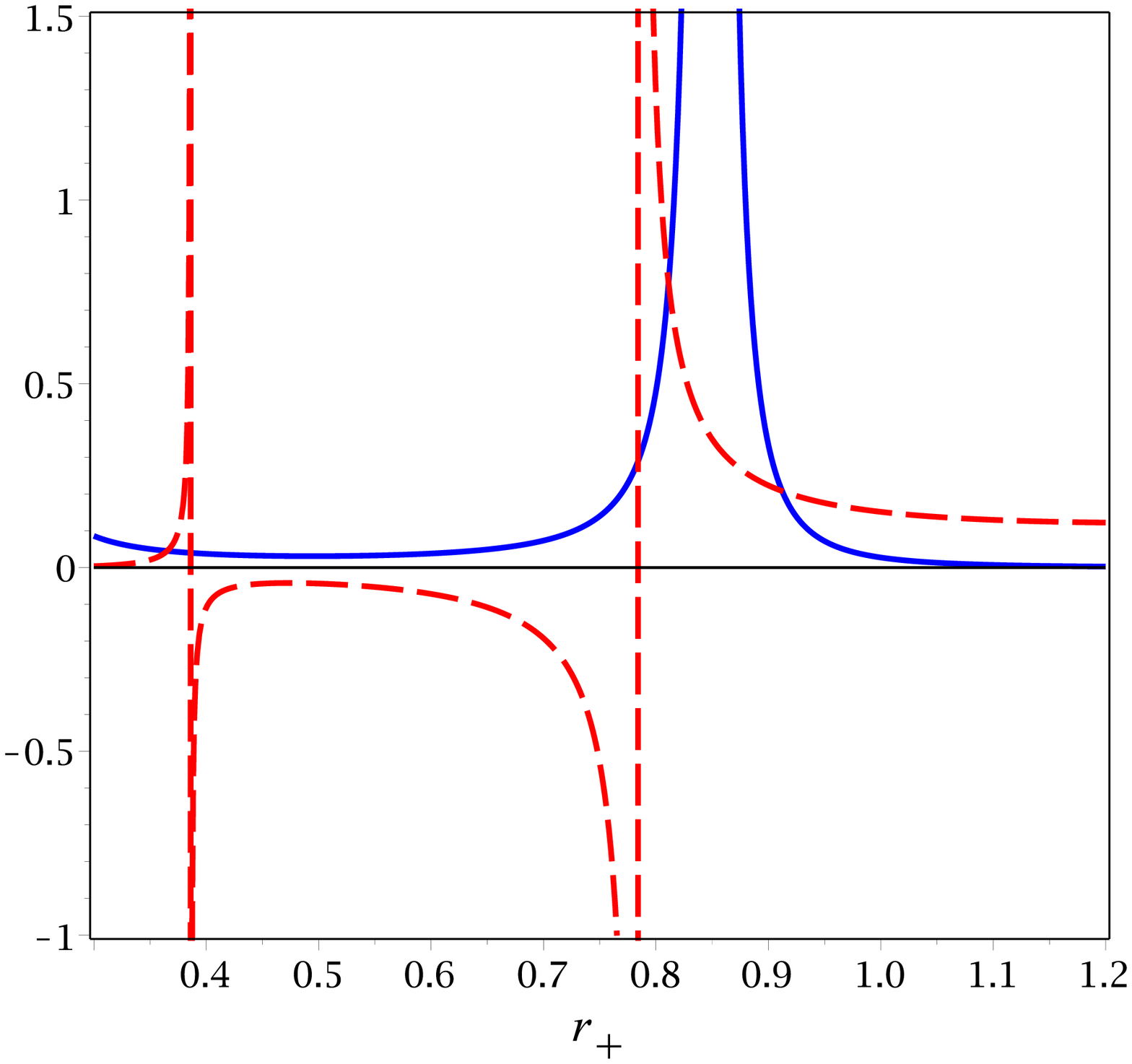}
\end{array}%
$%
\end{center}
\caption{Weinhold's Ricci scalar (continuous line) and heat
capacity (dashed line) versus $r_{+}$ for $q=0.1$, $\omega=0.2$,
$c_0=1$, $c_1=0.1$ and $P<P_C$ (different scales).} \label{Wfig}
\end{figure}

The Weinhold metric is given by
\begin{equation} \label{wm}
d s_{w}^2 =Mg_{ab}^{w}d x^{a}d x^{b},
\end{equation}
where $M$ is the mass, $g_{ab}^{w}=\frac{\partial^2
M(X^{c})}{\partial X^a \partial X^b}$ in which $X^a\equiv
X^a(S,N^i)$ and $N^i$'s are other extensive parameters. The
quantity that we care about is the Ricci scalar and its
divergences. Since the Ricci scalar's numerator is a finite smooth
function, we just care about its denominator to get the divergence
points and compare them with the divergence points and root of the
heat capacity. The Weinhold Ricci scalar denominator is
\begin{equation} \label{wmricci}
denom(R_W) =M^3 \left(M_{SS}M_{QQ}-M_{SQ}^2 \right)^2,
\end{equation}
where $M_X=\frac{\partial M}{\partial X}$,
$M_{XY}=\frac{\partial^2 M}{\partial X \partial Y}$ and
$M_{XX}=\frac{\partial^2 M}{\partial X^2}$. Considering Eq.
(\ref{wmricci}), we find that there is at least one divergence
point where $M_{SS}M_{QQ}=M_{SQ}^2$, and it is obvious from Fig.
\ref{Wfig}, that this point doesn't coincide with any of the
divergences or root of the heat capacity.

The Ruppeiner metric is defined as
\begin{equation} \label{rm}
d s_{R}^2 =-MT^{-1}g_{ab}^{w}d x^{a}d x^{b},
\end{equation}
which can be seen that, it is conformally related to Weinhold
metric. Again, we just care about the divergences of the Ruppeiner
Ricci scalar, so its denominator is our center of attention, since
its numerator is a finite smooth function
\begin{equation} \label{rmricci}
denom(R_R) =M T^3 \left(M_{SS}M_{QQ}-M_{SQ}^2 \right)^2.
\end{equation}

\begin{figure}[h!]
\begin{center}
$%
\begin{array}{cc}
\includegraphics[width=65 mm]{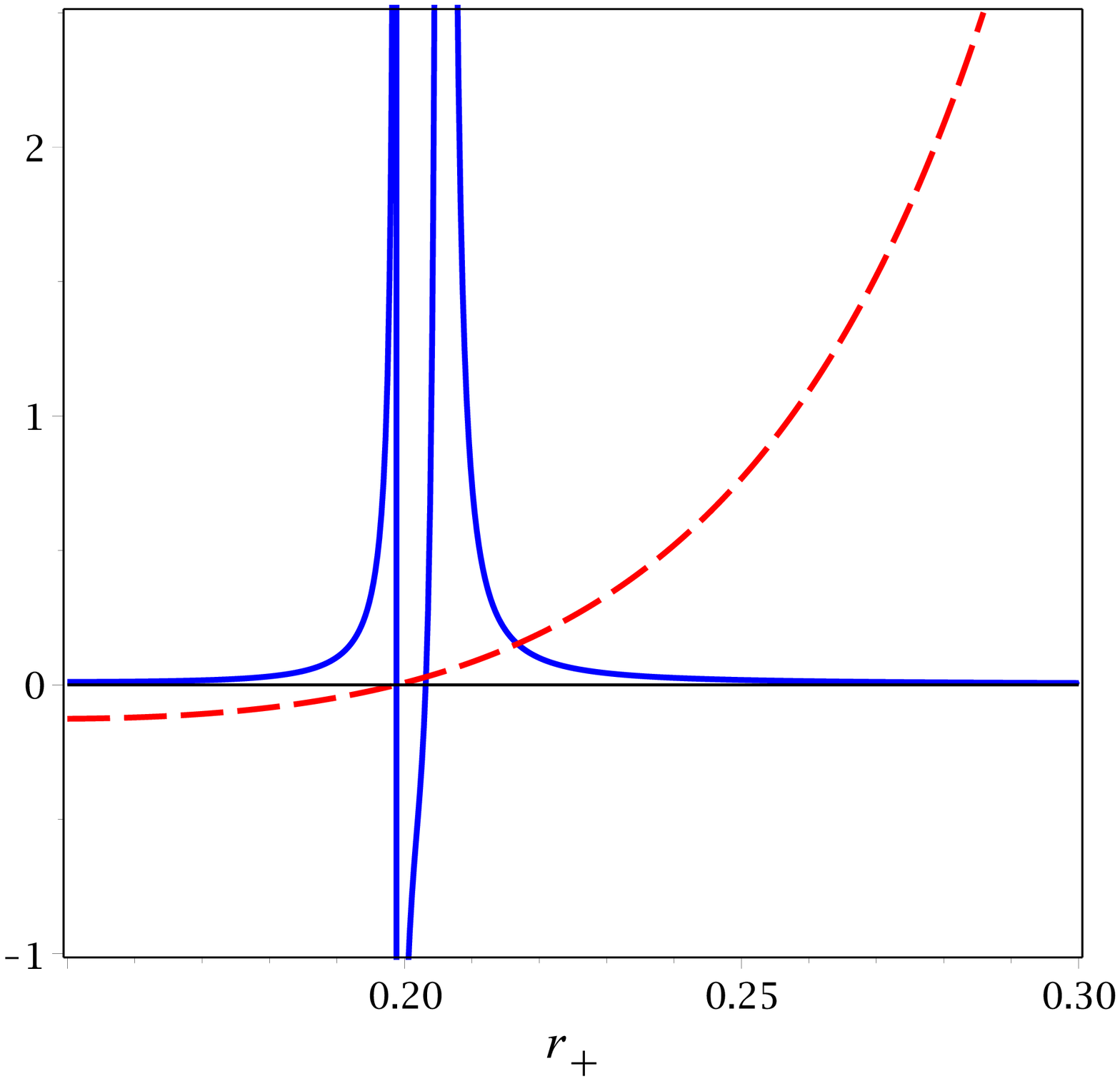} &
\includegraphics[width=69 mm]{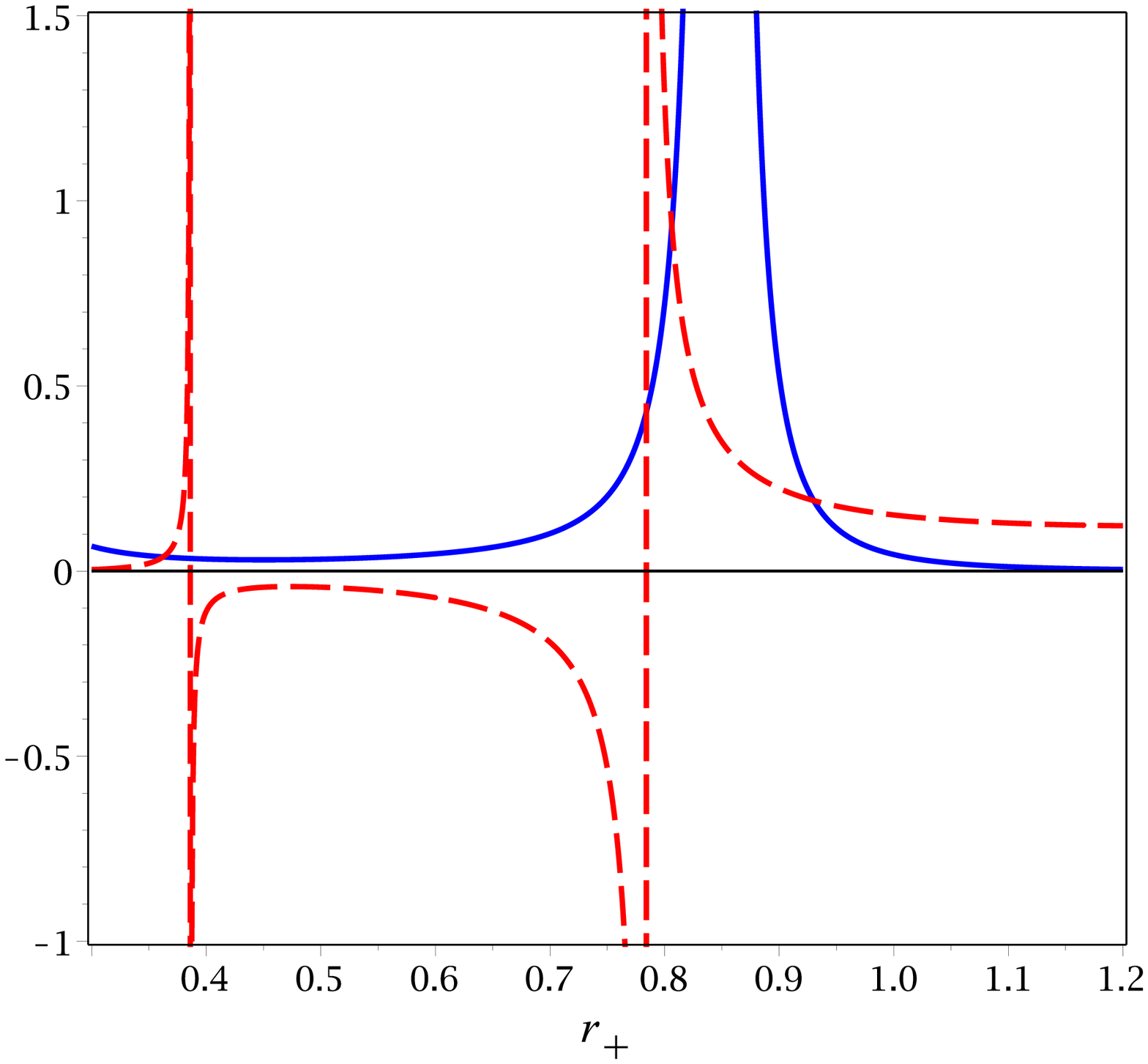}
\end{array}%
$%
\end{center}
\caption{Ruppeiner's Ricci scalar (continuous line) and heat
capacity (dashed line) versus $r_{+}$ for $q=0.1$, $\omega=0.2$,
$c_0=1$, $c_1=0.1$ and $P<P_C$ (different scales).} \label{Rfig}
\end{figure}

As it is obvious from Fig. \ref{Rfig}, one can find that the
divergence points of the Ruppeiner Ricci scalar don't match with
any of the divergences or root of the heat capacity. It can be
concluded from Figs. \ref{Wfig} and \ref{Rfig} that the phase
transition points reported by the Weinhold and Ruppeiner metrics
are not consistent with phase transition points indicated by the
heat capacity.

\begin{figure}[h!]
\begin{center}
$%
\begin{array}{cc}
\includegraphics[width=68 mm]{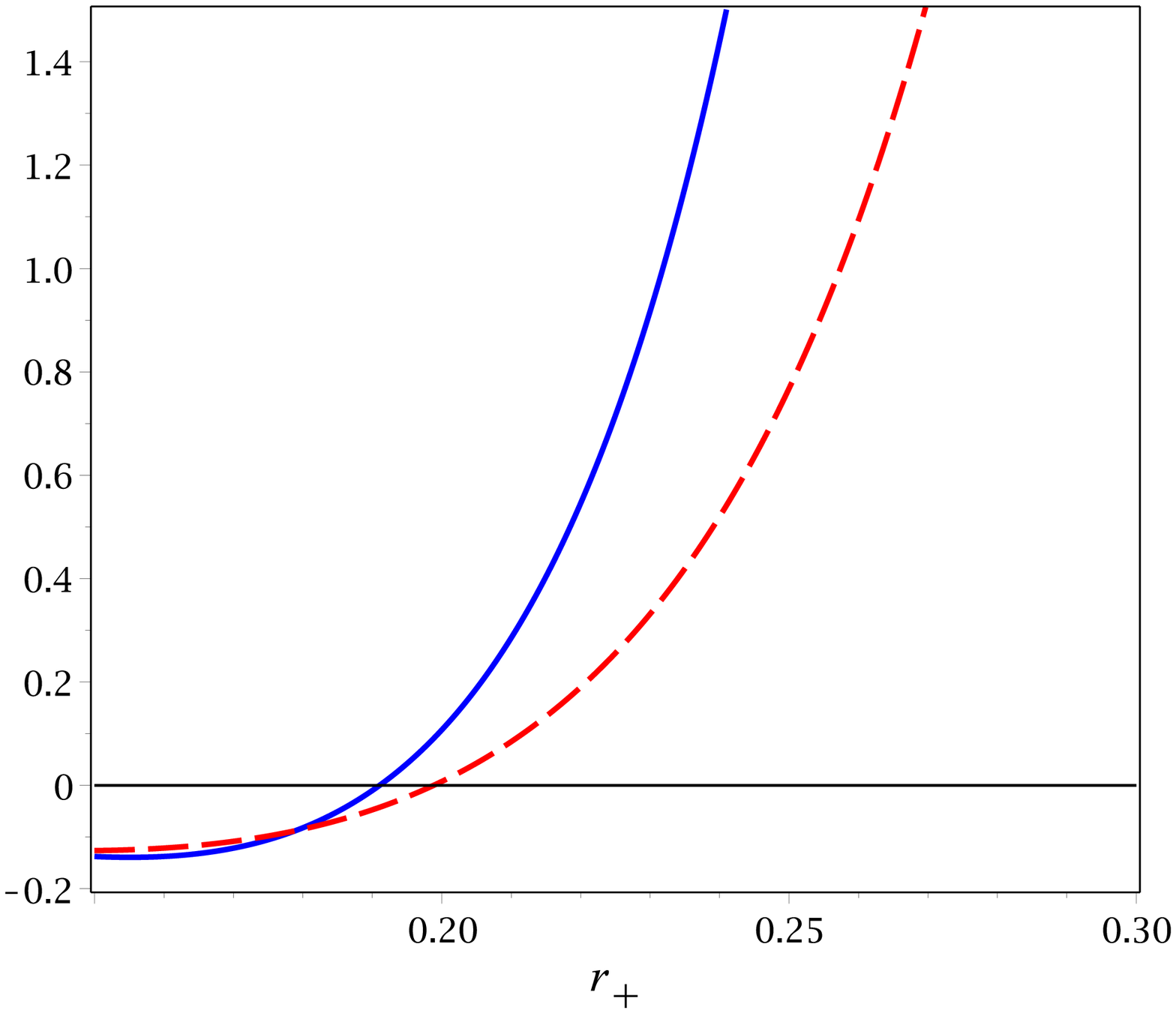} &
\includegraphics[width=65 mm]{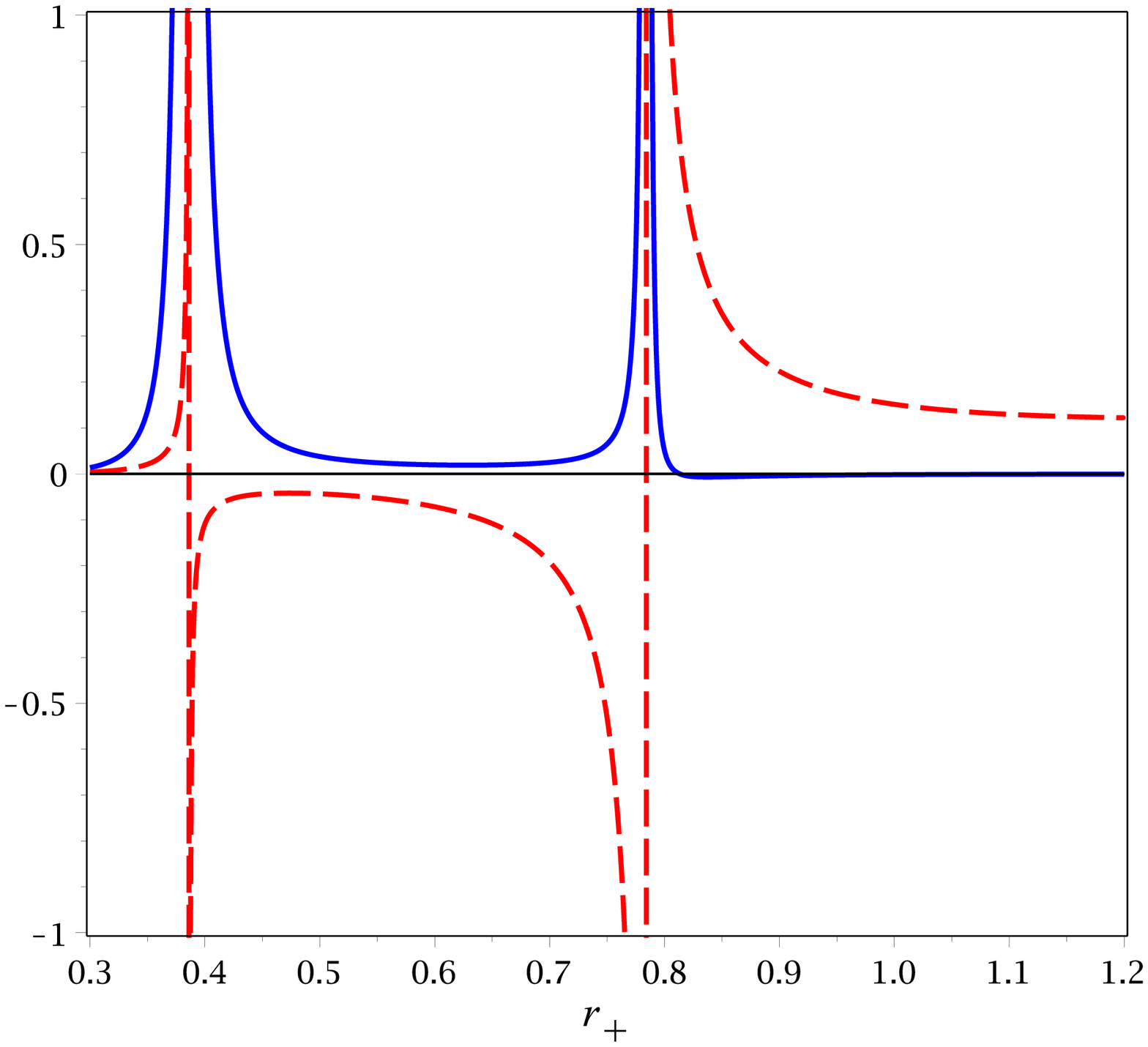}
\end{array}%
$%
\end{center}
\caption{Quevedo's Ricci scalar (continuous line) and heat
capacity (dashed line) versus $r_{+}$ for $q=0.1$, $\omega=0.2$,
$c_0=1$, $c_1=0.1$ and $P<P_C$ (different scales).} \label{Qfig}
\end{figure}

\begin{figure}[h!]
\begin{center}
$%
\begin{array}{c}
\includegraphics[width=65 mm]{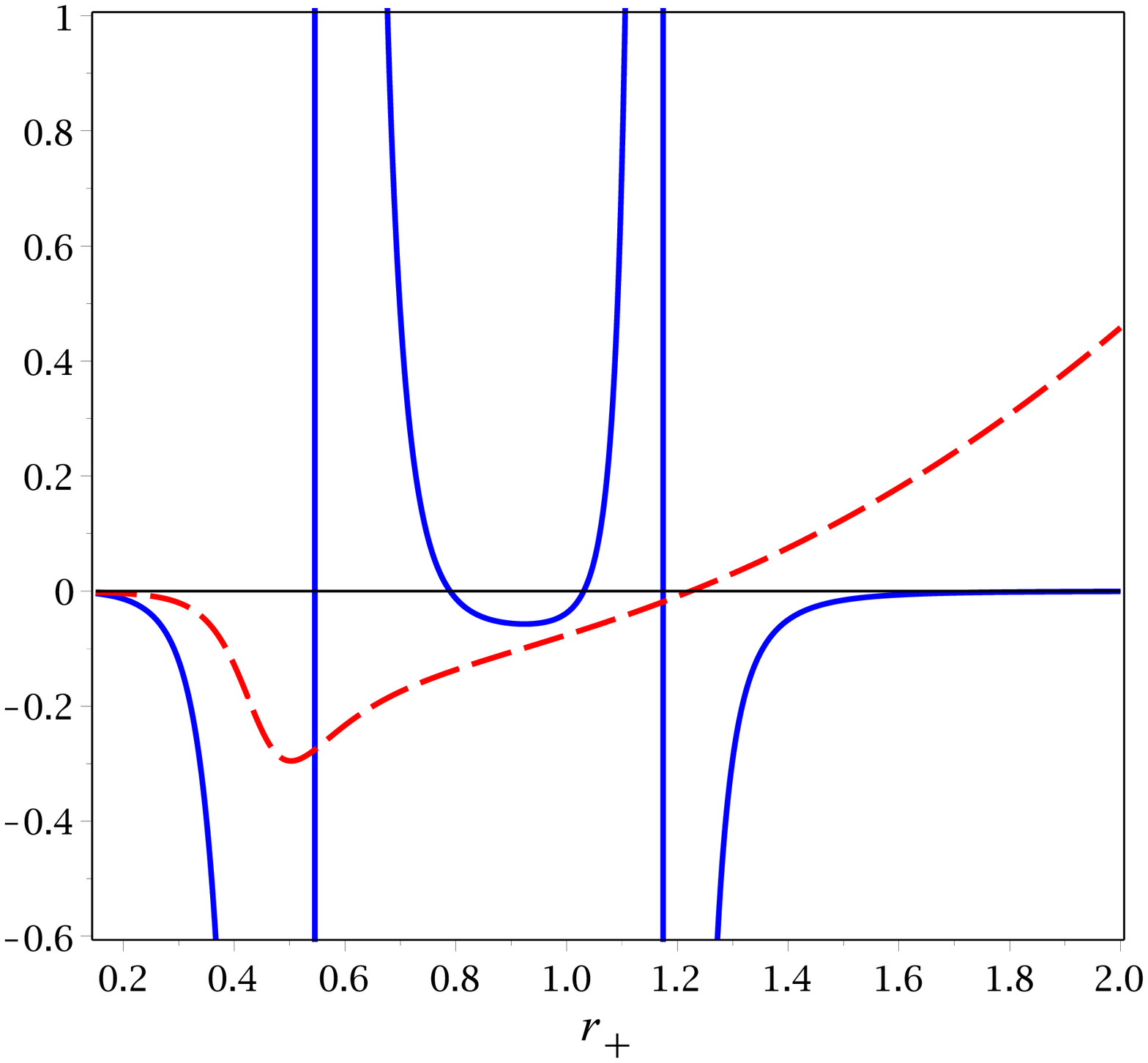}
\end{array}%
$%
\end{center}
\caption{Quevedo's Ricci scalar (continuous line) and heat
capacity (dashed line) versus $r_{+}$ for $q=0.1$, $\omega=0.2$,
$c_0=1$, $c_1=2$ and $P>P_C$ (mismatches are clear).} \label{Anom}
\end{figure}

Next, we examine the Quevedo metric which is defined as
\begin{equation} \label{qm}
ds_{Q}^2 =\left(SM_S+QM_Q \right) \left(-M_{SS}dS^2+M_{QQ}dQ^2
\right).
\end{equation}
Like before for getting the divergences of the Ricci scalar, we
isolate its denominator which is the form
\begin{equation} \label{qmricci}
denom(R_Q)= M_{SS}^2M_{QQ}^2 \left(SM_S+QM_Q \right)^3.
\end{equation}
Existence of $M_{SS}$ in the denominator ensures the consistence
of the phase transition points of Ricci scalar with the
corresponding points of heat capacity (see right panel of Fig.
\ref{Qfig}), but as it is shown in Fig. \ref{Anom}, one can see
that there is another divergence point (related to the root of
$SM_S+QM_Q$) that is not consistent with the results of heat
capacity. In addition, left panel of Fig. \ref{Qfig} confirms that
Quevedo's Ricci scalar could not characterize the root of heat
capacity.

\begin{figure}[h!]
\begin{center}
$%
\begin{array}{cc}
\includegraphics[width=65 mm]{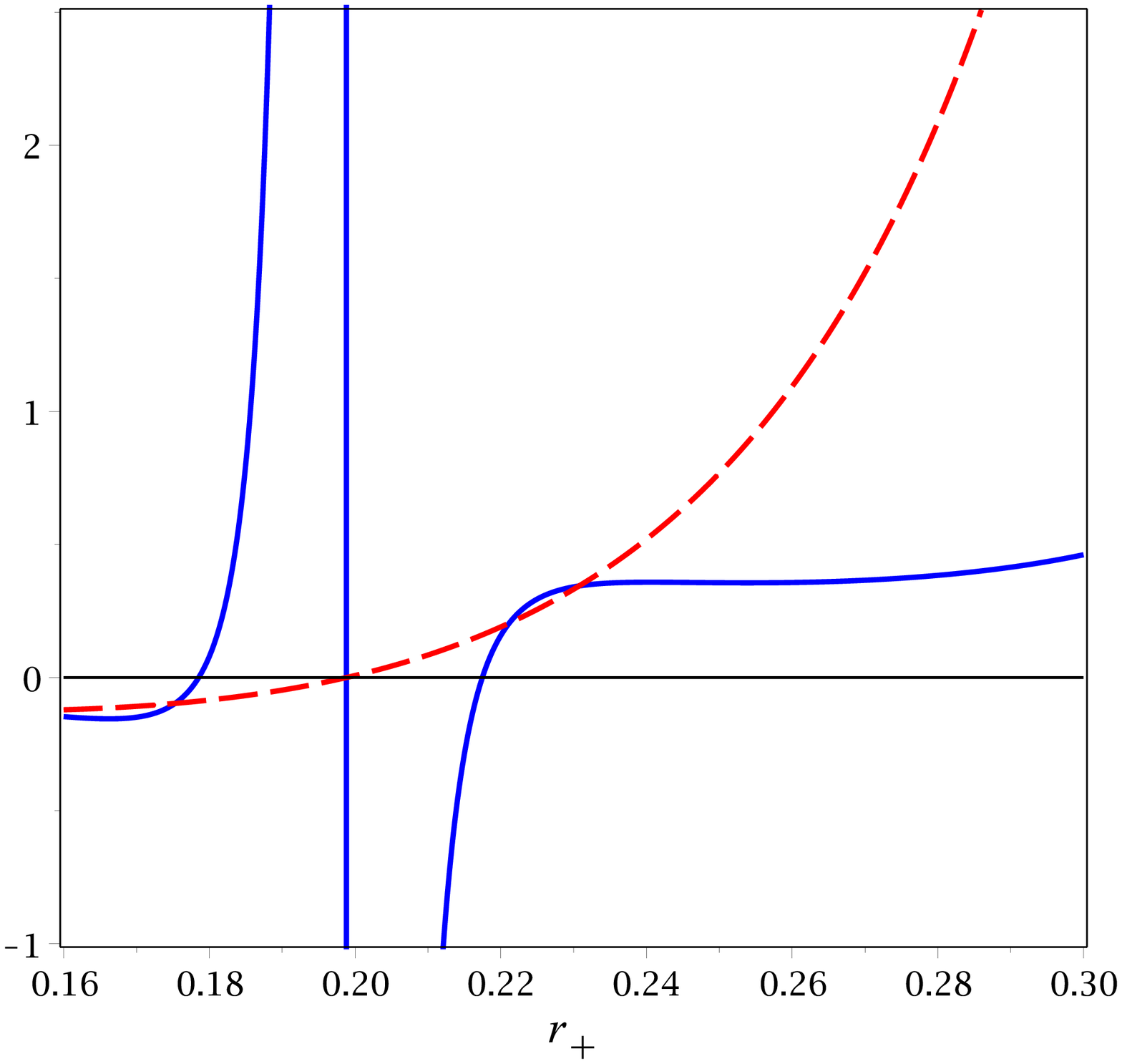} &
\includegraphics[width=68 mm]{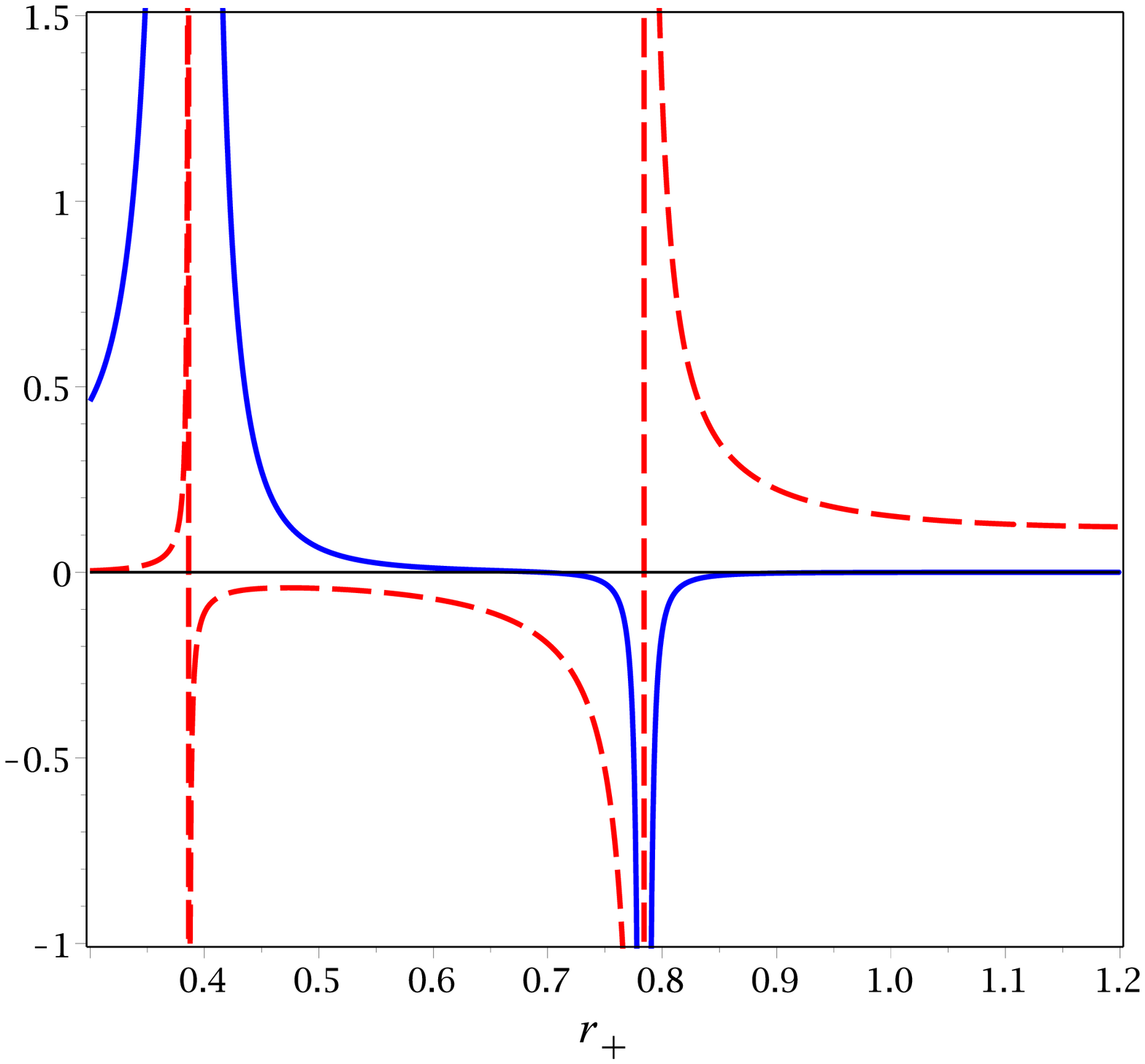}
\end{array}%
$%
\end{center}
\caption{HPEM's Ricci scalar (continuous line) and heat capacity
(dashed line) versus $r_{+}$ for $q=0.1$, $\omega=0.2$, $c_0=1$,
$c_1=0.1$ and $P<P_C$ (different scales).} \label{HPEMfig}
\end{figure}

Finally, we employ the HPEM metric which has the form of
\begin{equation} \label{hpemm}
ds_{HPEM}^2 =(SM_S/M_{QQ}^3)(-M_{SS}dS^2+M_{QQ}dQ^2),
\end{equation}
and the denominator of its Ricci scalar can be simplified as
\begin{equation} \label{hpemmricci}
denom(R_HPEM)= 2S^3M_{SS}^2M_s^3.
\end{equation}
As it is evident from Eq. (\ref{hpemmricci}), we can see that
there exist two sets of divergence points, one corresponding to
$M_S=0$ and the other one corresponds to $M_{SS}=0$, resulting in
consistence of it divergence points with both the root and phase
transition points of the heat capacity. One interesting feature
which can be seen from graphs in Fig. \ref{HPEMfig} is the
behavior of HPEM metric Ricci scalar in the vicinity of its
divergence points, meaning that the divergence points of Ricci
scalar related to the roots of heat capacity is distinguishable
from its divergence points related to divergences of heat capacity
based on the Ricci scalar behavior.


\section{Thermal Fluctuations}

We will now discuss the effects of thermal fluctuations on the
thermodynamics and stability of this black string solution. As
this solution can be an AdS solution, it has to have a CFT dual,
and this CFT can be used to obtain the microstates of this system
\cite{cjp0}. So,  let us assume that these   microstates, which
give rise to the entropy of this  black string, are denoted by
$\rho$. Now it is possible to write the partition function using
such microstates $\rho$ as
\begin{equation}  \label{S1}
Z=\int{dE\rho e^{-{E}/{T}}},
\end{equation}
where  $E$ is internal energy, which is related to
Helmholtz free energy $F$ as $E=ST+F$. It may be noted that the
Helmholtz free energy $F$  can be written as
\begin{equation}  \label{S2}
F=-\int{TdS}.
\end{equation}
Now using the relations (\ref{Entropy}) and (\ref{Temp}), one can obtain,
\begin{equation}  \label{S3}
F=\left\vert 1-\omega_{0}^{2}\right\vert\left(\frac{\Lambda}{12}r_{+}^{3}+%
\frac{c_{1}}{4}r_{+}^{2}-\frac{c_{0}}{4}r_{+}-\frac{q^{2}}{r_{+}}\right).
\end{equation}
Hence, the partition function of canonical ensemble for this black
string can be constructed  using these microstates. Having
partition function for black strings, we can  use the first order
Taylor expansion around the equilibrium to express these density
of states as  \cite{cjp0, CJP},
\begin{equation}  \label{S4}
\rho=\frac{e^{S}}{\sqrt{2\pi \frac{d^{2}S}{d\beta_{k}^{2}}}},
\end{equation}
where $\beta_{k}=\frac{1}{T}$ is the equilibrium  temperature. Now  we can use the statistical
relation $\bar{S}=\ln{\rho}$ to obtain corrected entropy for black strings     \cite{CJP, cjp1, cjp2, cjp4}
\begin{equation}  \label{S5}
\bar{S}=S-\frac{1}{2}\gamma\ln{\left(2\pi\left[T^{4}\frac{\partial^{2}S}{%
\partial T^{2}}+2T^{3}\frac{\partial S}{\partial T}\right]\right)},
\end{equation}
where $S$ is given by the equation (\ref{Entropy}). As the
coefficient of the leading order correction term to the black hole
entropy depends on the model used, we have added a free parameter
$\gamma$ to measure the strength of this correction term
\cite{EPL}. It may be noted that  we can  set $\gamma=1$ for the
corrected thermodynamics. Furthermore, the corrected
thermodynamics   reduces  to the ordinary thermodynamics at
$\gamma=0$. It has been demonstrated that these thermal
fluctuation in the thermodynamics occurs due to quantum
fluctuation of the metric, and becomes important when the black
hole is small \cite{NPB}.   Now, we would like to study effect of
such corrections on the thermodynamics of rotating black string in
massive gravity.

In this case, we can obtain corrected Helmholtz free energy,
\begin{equation}  \label{S6}
\bar{F}=-\int{Td\bar{S}}.
\end{equation}
Using the corrected entropy (\ref{S5}),  we can find,
\begin{equation}  \label{S7}
\bar{F}=F+\frac{\gamma}{2}\int{\frac{T^{2}S_{3}+6TS_{2}+6S_{1}} {%
TS_{2}+2S_{1}}dT},
\end{equation}
where,  we have
\begin{eqnarray}  \label{S8}
S_{1}\equiv\frac{\partial S}{\partial T}&=&-\frac{4\pi^{2}r_{+}^{5}}{%
\left\vert 1-\omega_{0}^{2}\right\vert (\Lambda
r_{+}^{4}-12q^{2}+c_{0}r_{+}^{2})},  \notag \\
S_{2}\equiv\frac{\partial^{2}S}{\partial T^{2}}&=&\frac{16\pi^{3}r_{+}^{8}(%
\Lambda r_{+}^{4}-60l^{2}q^{2}+3c_{0}r_{+}^{2})}{\left\vert
1-\omega_{0}^{2}\right\vert^{2} (\Lambda
r_{+}^{4}-12q^{2}+c_{0}r_{+}^{2})^{3}}  \notag \\
S_{3}\equiv\frac{\partial^{3}S}{\partial T^{3}}&=&-\frac{768%
\pi^{4}r_{+}^{11}((8\Lambda q^{2}+c_{0}^{2})
r_{+}^{4}-40c_{0}q^{2}r_{+}^{2}+480q^{4})}{\left\vert
1-\omega_{0}^{2}\right\vert^{3} (\Lambda
r_{+}^{4}-12q^{2}+c_{0}r_{+}^{2})^{5}}.
\end{eqnarray}

\begin{figure}[h!]
 \begin{center}$
 \begin{array}{cccc}
\includegraphics[width=51 mm]{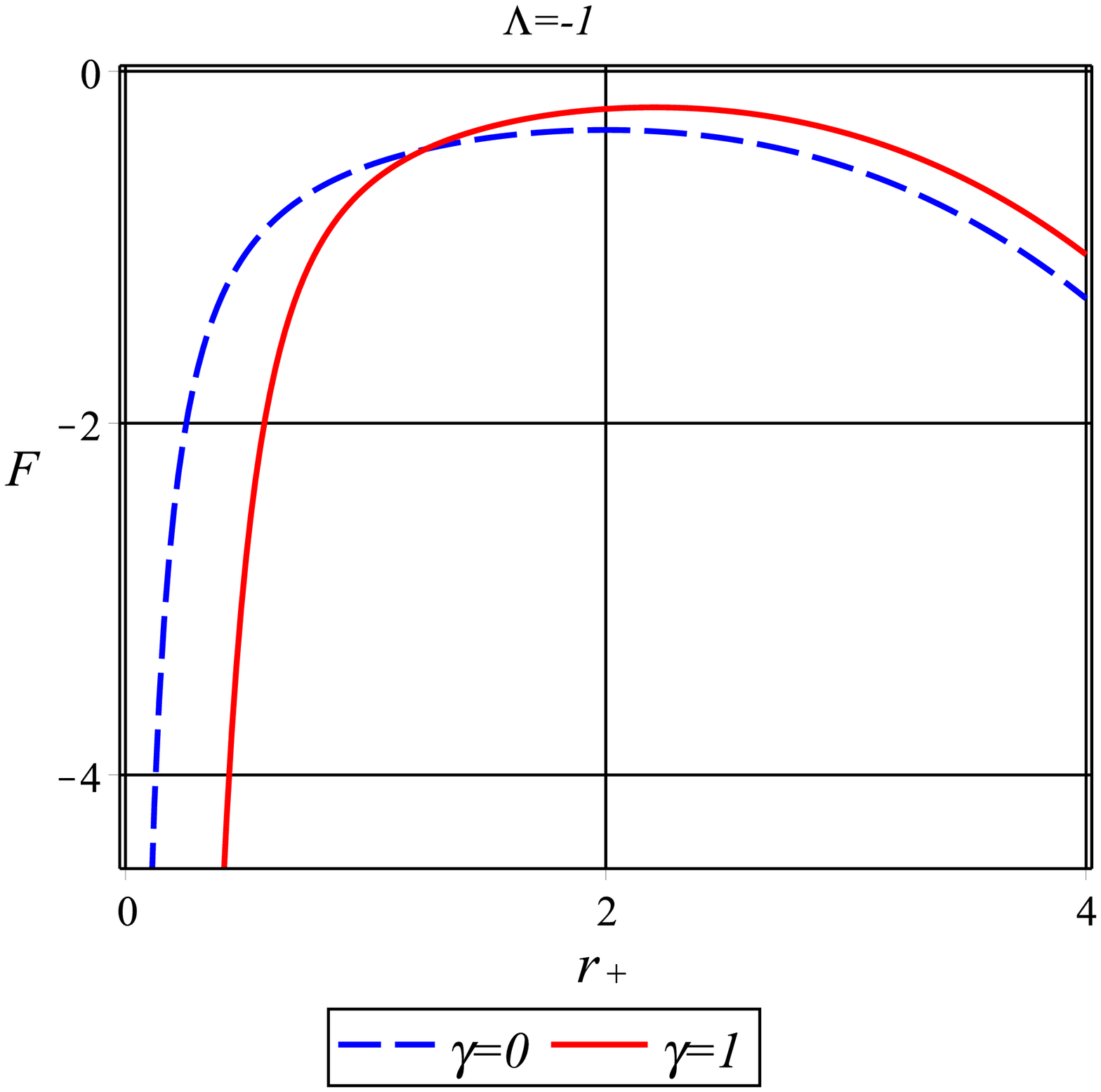}\includegraphics[width=51 mm]{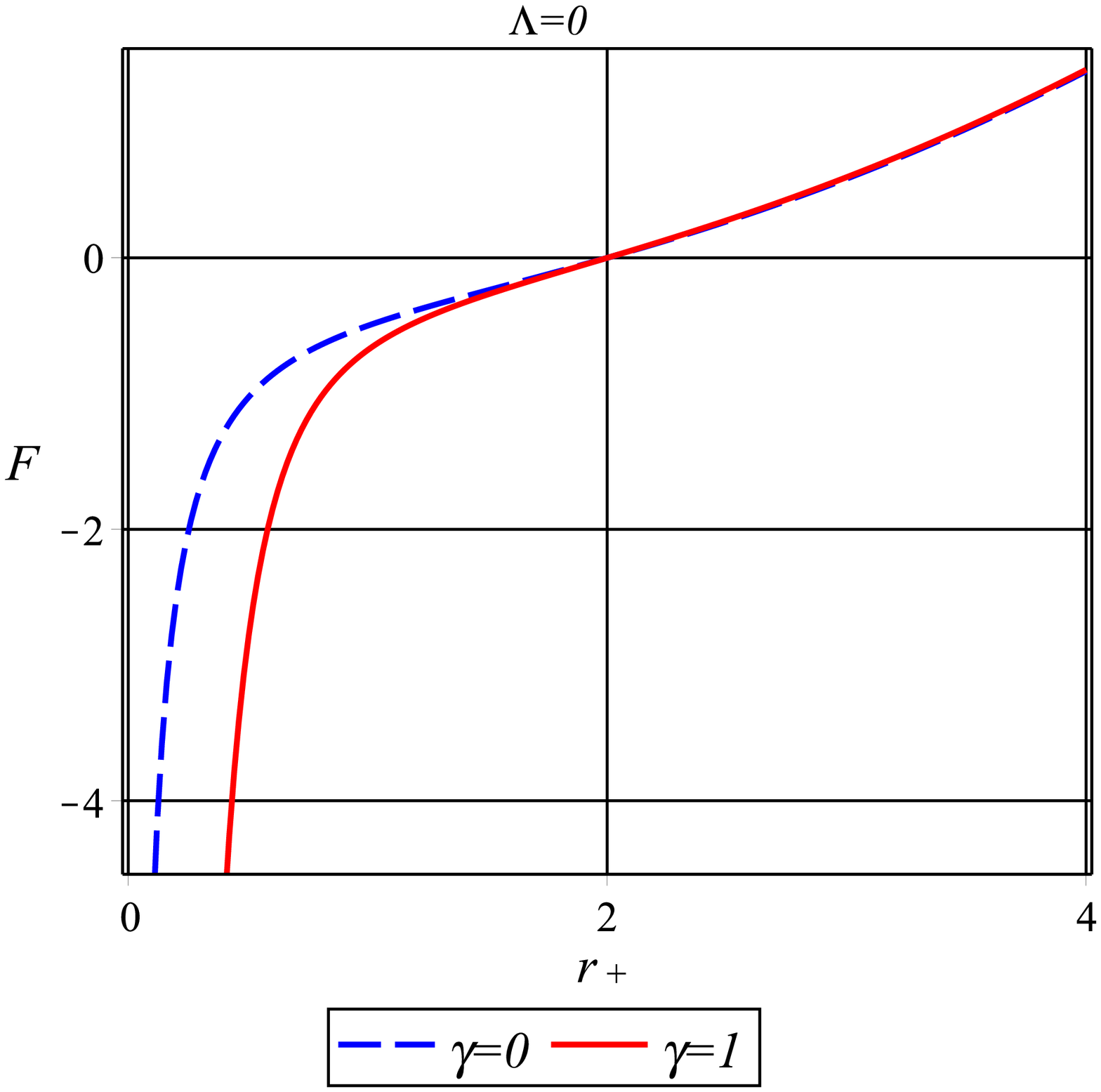}\includegraphics[width=51 mm]{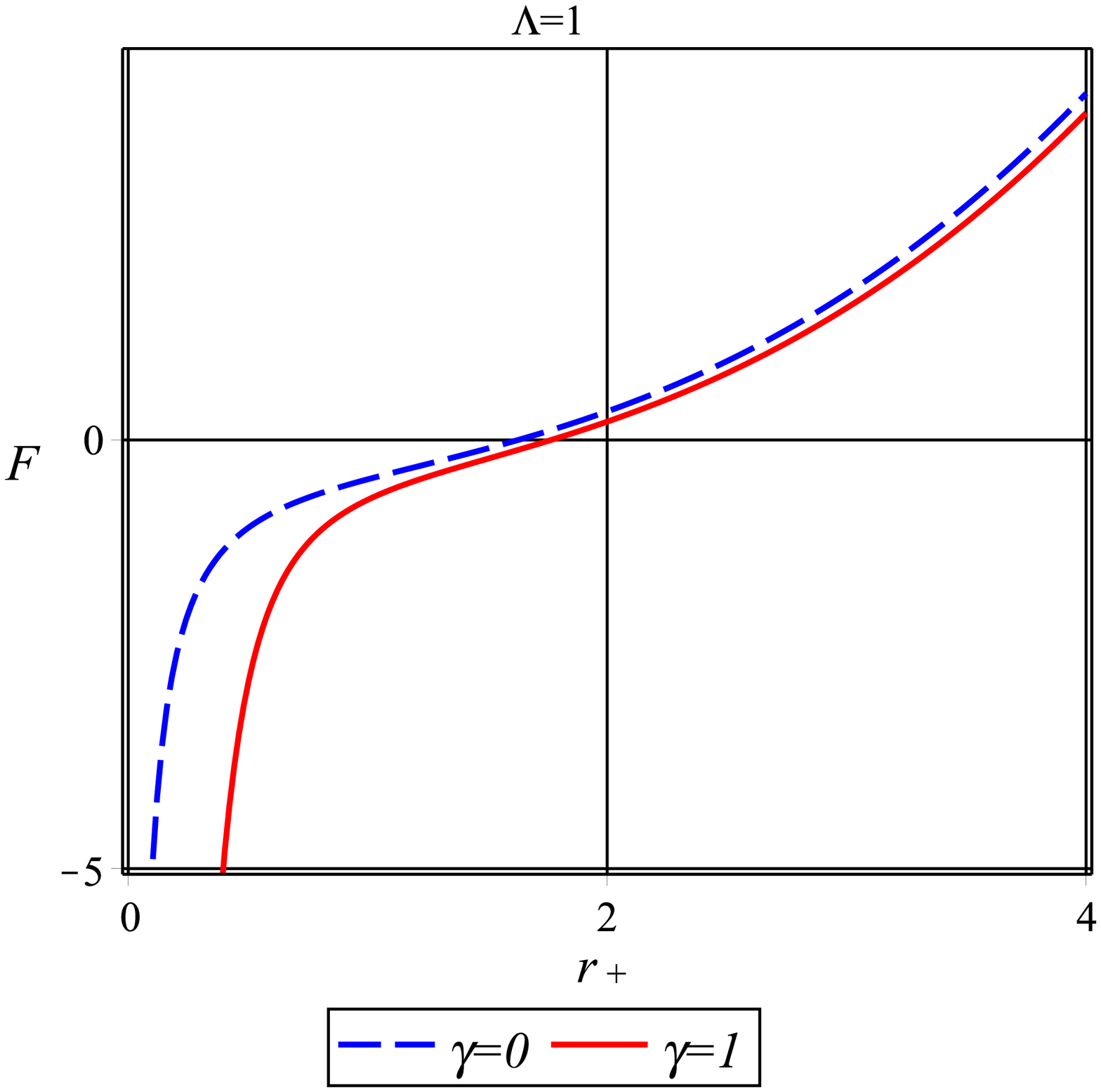}
 \end{array}$
 \end{center}
\caption{Corrected Helmholtz free energy $\bar{F}$ in terms of horizon
radius for $c_{0}=c_{1}=l=q=1$, and $\protect\omega^{2}=0.5$.}
 \label{F}
\end{figure}

Now as the  thermal fluctuations are important, when the black hole
 is small, ($r_{+}\ll1$, hence $S_{3}\ll S_{2}\ll S_{1}$), otherwise $%
\gamma\ll 1$ and $\bar{F}\approx F$. Therefore, neglecting $S_{3}$ and $%
S_{2} $, and keeping only $S_{1}$, we can obtain,
\begin{equation}  \label{S9}
\bar{F}\approx F+\frac{3\gamma}{8}\left\vert 1-\omega_{0}^{2}\right\vert%
\frac{-\Lambda r_{+}^{4}-4q^{2}+c_{0}r_{+}^{2}}{\pi r_{+}^{3}}.
\end{equation}
In the plots of the Fig. \ref{F}, we can see typical behavior of
the Helmholtz free energy for different values of $\Lambda$.
Dashed blue lines show original $F$, given by the equation
(\ref{S3}), while solid red line represent the  corrected
quantity. In the cases of $\Lambda\geq0$, the effect of thermal
fluctuations are important for smaller $r_{+}$, hence the
corrections can be neglected for the large black hole. In the case
of $\Lambda<0$, there is a critical radius, where the effect of
thermal fluctuations vanishes. Before the critical radius the
thermal fluctuations decrease the energy, and  after the critical
radius, we observe that the Helmholtz free energy increases.

\begin{figure}[h!]
 \begin{center}$
 \begin{array}{cccc}
\includegraphics[width=52 mm]{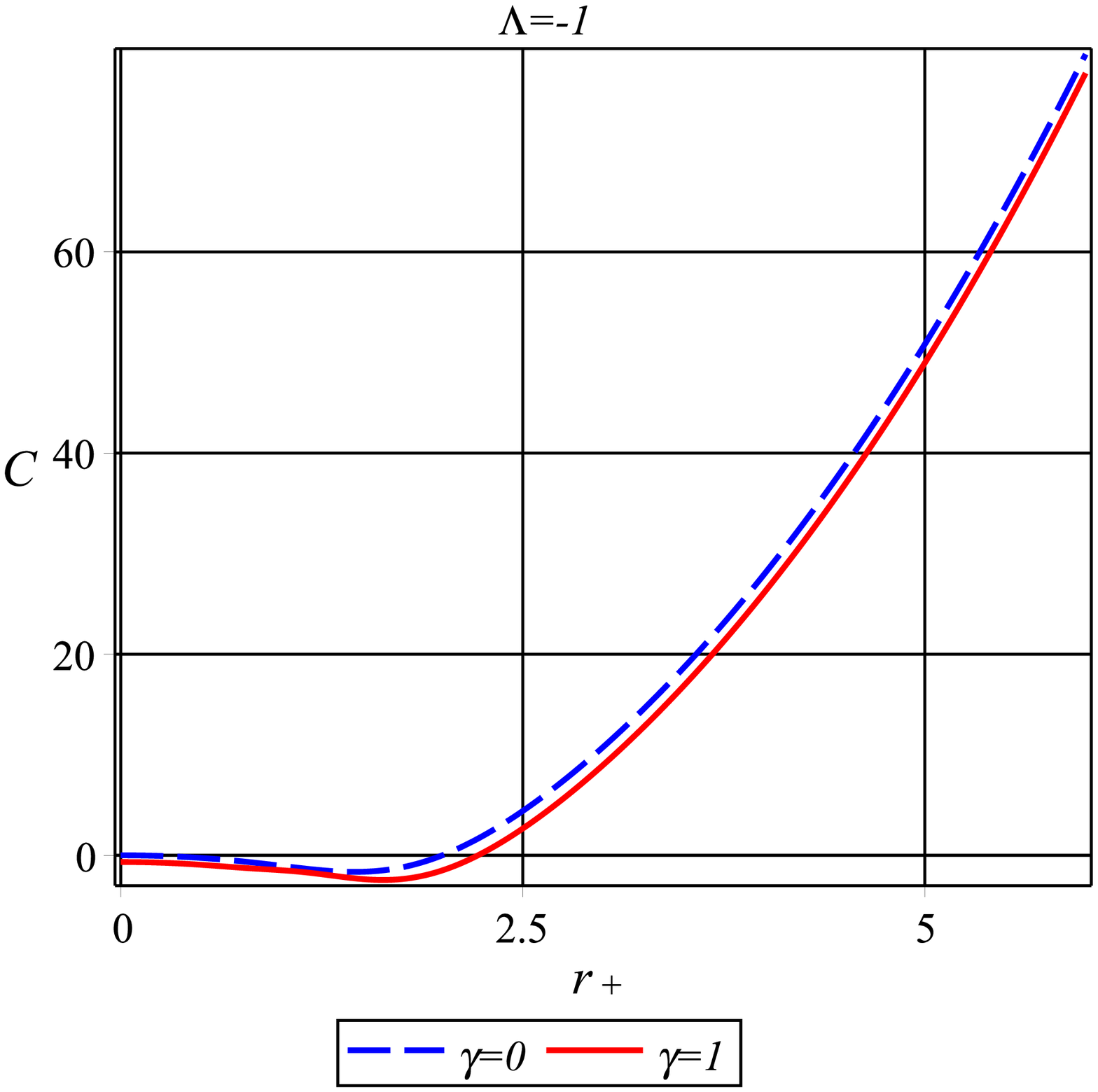}\includegraphics[width=52 mm]{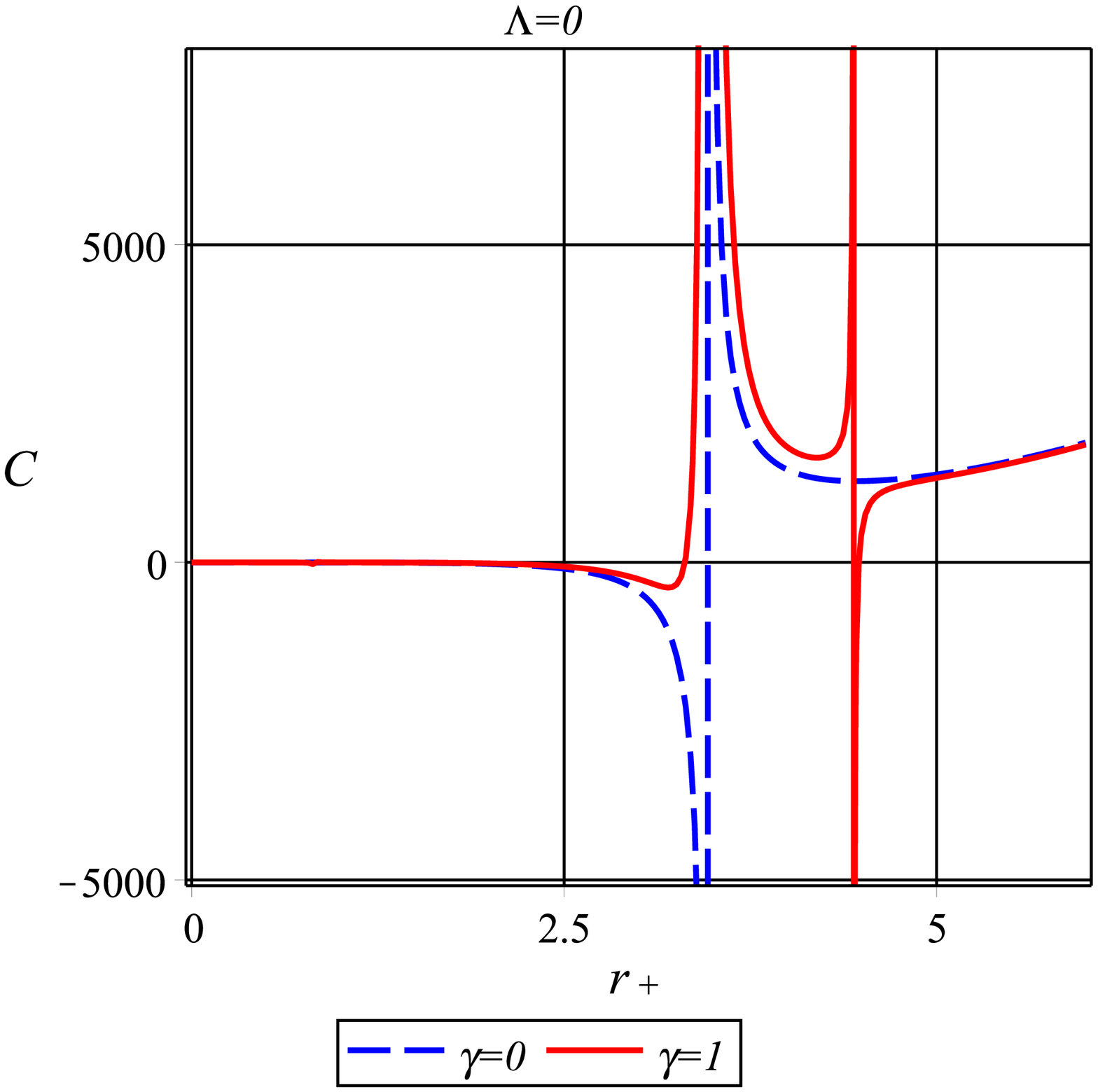}\includegraphics[width=52 mm]{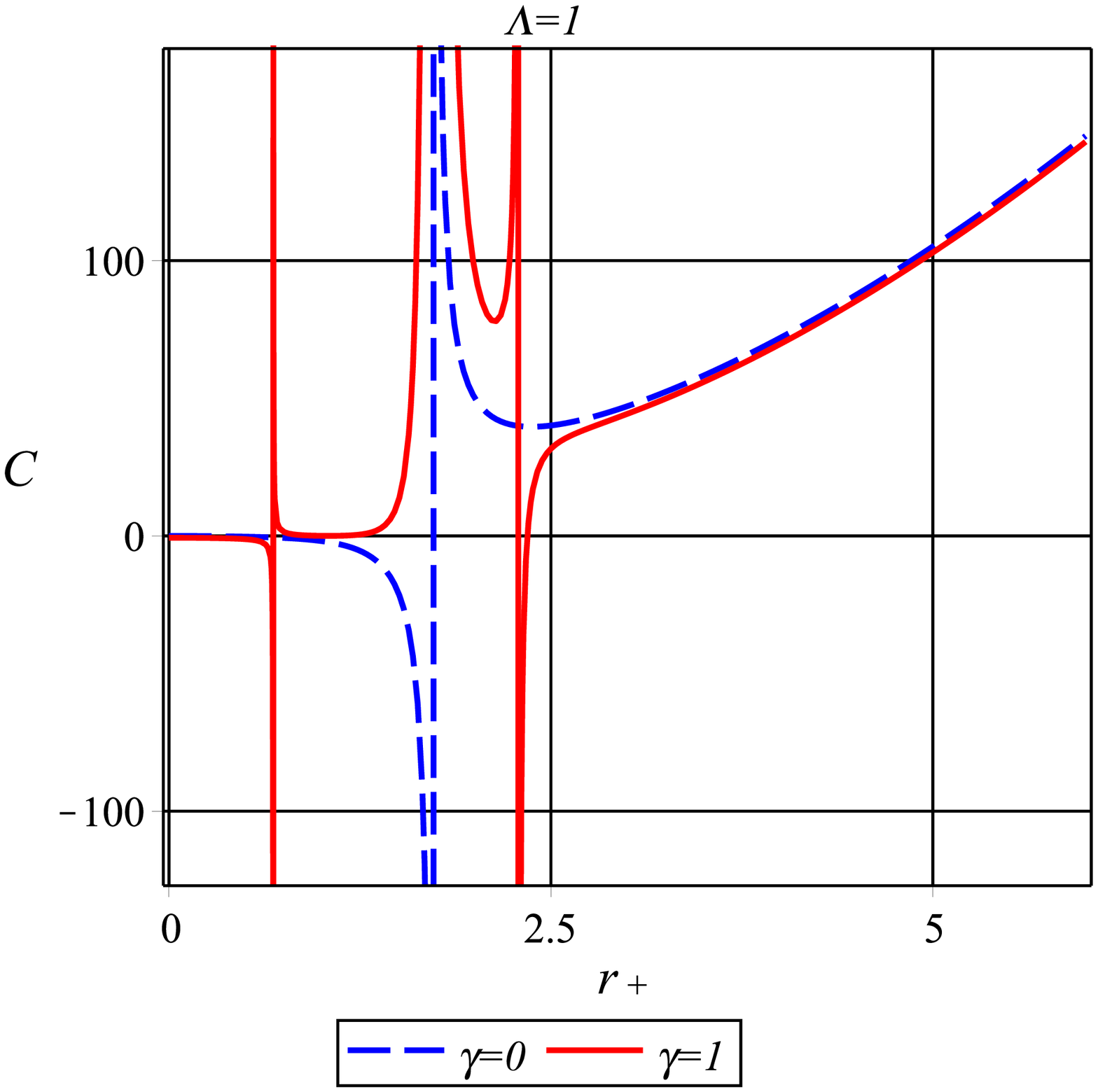}
 \end{array}$
 \end{center}
\caption{Corrected specific heat $\bar{C}$ in terms of horizon radius for $%
c_{0}=c_{1}=l=q=1$, and $\protect\omega^{2}=0.5$.}
 \label{C}
\end{figure}

We can also calculate the  corrected specific heat using
\begin{equation}  \label{S10}
\bar{C}=T\frac{d\bar{S}}{dT}=C+C(\gamma),
\end{equation}
where $C(\gamma)$ is the correction term,  which depend on $\gamma$,  and
\begin{equation}  \label{S11}
C=\frac{\pi r_{+}^{2}(\Lambda
r_{+}^{4}-2c_{1}r_{+}^{3}+4q^{2}-c_{0}r_{+}^{2})}{(\Lambda
r_{+}^{4}-12q^{2}+c_{0}r_{+}^{2})},
\end{equation}
in agreement with the equation (\ref{CQ}). In the plots of the
Fig. \ref{C}, we can observe the typical behavior of the specific
heat for different values of $\Lambda$. Dashed blue lines show
ordinary $C$ given in the Eq.  (\ref{S11}), while solid red line
represent the  corrected quantity obtained in Eq. (\ref{S10}). As
before,   $\Lambda=-1$ is different from other values of
$\Lambda\geq0$. In the case of $\Lambda=-1$, the  thermal
fluctuations decrease the specific heat (see the left plot of the
Fig. \ref{C}). In all cases, we can observe some instabilities at
small $r_{+}$. We can interpret it as the black string  becoming
small,  due to the hawking radiation, and  then becoming
unstable.   So, there   seems to be  a minimum size for the stable
black strings in massive gravity.

Now for  $\Lambda\geq0$, with  thermal fluctuations, we observe
that there is a  second order phase transition (see the right plot
of the Fig. \ref{C}). As expected, for the large $r_{+}$,  there
is no important differences between ordinary and corrected
quantities. The the  thermal fluctuations modify the  stability
and phase transition of a charged rotating black string in massive
gravity.

\begin{figure}
\begin{center}$
\begin{array}{cccc}
\includegraphics[width=75 mm]{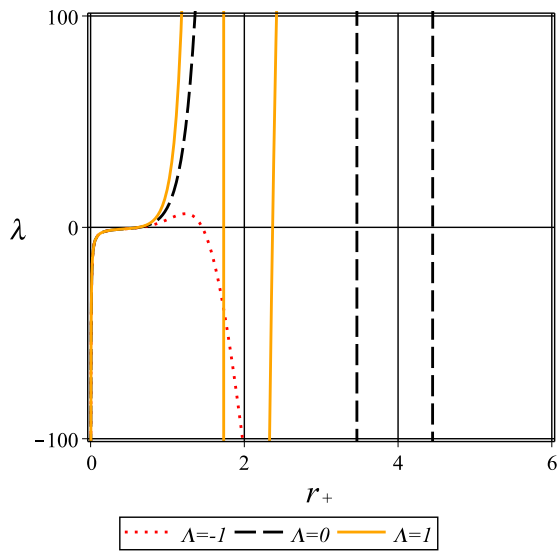}
\end{array}$
\end{center}
\caption{Trace of Hessian matrix in terms of horizon radius for $c_{0}=c_{1}=l=q=Q=\alpha_{g}=\gamma=1$, and $\omega^{2}=0.5$.}
\label{Hess}
\end{figure}

It is argued that in the presence of electric charge, the heat
capacity analysis is not sufficient to analyze  the stability of
system \cite{Hess1, Hess2, Hess3}. In such a  system, we need the
entire Hessian matrix of the Helmholtz free energy. This is
constructed by second derivatives of Helmholtz free energy with
respect to temperature and electric potential. So, the  Hessian
metric for the black string can be written using
\begin{eqnarray}\label{Hess1}
H_{11}\equiv\frac{\partial^{2}\bar{F}}{\partial T^{2}}&=&-\frac{8\pi^{2}r_{+}^{5}}{\left\vert 1-\omega_{0}^{2}\right\vert (\Lambda r_{+}^{4}-12q^{2}+c_{0}r_{+}^{2})}X,\nonumber\\
H_{12}\equiv\frac{\partial^{2}\bar{F}}{\partial T\partial\Phi}&=&\frac{\alpha_{g}\pi r_{+}^{3}}{Q(\Lambda r_{+}^{4}-12q^{2}+c_{0}r_{+}^{2})^2}X,\nonumber\\
H_{21}\equiv\frac{\partial^{2}\bar{F}}{\partial \Phi\partial T}&=&-\frac{\alpha_{g} r_{+}}{2Q(\Lambda r_{+}^{4}-12q^{2}+c_{0}r_{+}^{2})}Y,\nonumber\\
H_{22}\equiv\frac{\partial^{2}\bar{F}}{\partial \Phi^{2}}&=&-\frac{\left\vert 1-\omega_{0}^{2}\right\vert\alpha_{g}^{2}}{16Q^{2}r_{+}}Y,
\end{eqnarray}
where
\begin{eqnarray}\label{Hess2}
X&=&\Lambda^{2}r_{+}^{8}+c_{1}\Lambda r_{+}+2c_{0}\Lambda r_{+}^{6}+3c_{0}c_{1}r_{+}^{5}\nonumber\\
&-&(40\Lambda q^{2}+c_{0}^{2})r_{+}^{4}-60c_{1}q^{2}r_{+}^{3}+24c_{0}q^{2}r_{+}^{2}-48q^{4},\nonumber\\
Y&=&-4\pi\Lambda r_{+}^{6}-6\pi c_{1}r_{+}^{5}+(3\gamma \Lambda+2\pi c_{0}) r_{+}^{4}+36\gamma q^{2}.
\end{eqnarray}
Now Hessian matrix $\mathcal{H}$ is given by,
\begin{eqnarray}
\mathcal{H}=\left(\begin{array}{ccc}
H_{11} & H_{12}\\
H_{21} & H_{22}\\
\end{array}\right).
\end{eqnarray}
It is easy to check that determinant of above matrix is zero.
Thus, one of its eigenvalues is zero.  The other is given by the
trace of matrix,
\begin{equation}\label{Hess3}
\lambda=H_{11}+H_{22}.
\end{equation}
In the plot of the Fig. \ref{Hess}, we draw $\lambda$, and observe
that the asymptotic points coincide with the Fig. \ref{C}. Hence
our discussion about the  stability are valid, and in agreement
with analysis done using the  specific heat of this system.

It may be noted that the  extended first law is valid even with
 these corrected entropy terms (\ref{S5}). In fact, the extended
first law be satisfied  for such correction  terms, as the
corrected entropy can be write  as \cite{cjp0, CJP}
\begin{equation}\label{S-last}
\bar{S}=S+\gamma_{1}S_{1}+\gamma_{2}S_{2}+\cdots,
\end{equation}
where $\gamma_{1}\equiv-\frac{1}{2}\gamma$ and $S_{1}\equiv\ln{\left(2\pi\left[T^{4}\frac{\partial^{2}S}{%
\partial T^{2}}+2T^{3}\frac{\partial S}{\partial T}\right]\right)}$.
In order to satisfy the extended first law, we should have,
\begin{equation}
\gamma_{1}dS_{1}+\gamma_{2}dS_{2}+\cdots=0.
\end{equation}
Above condition could be satisfied  by suitable choice of
coefficients $\gamma_{1}$, $\gamma_{2}....$,  as the Lagrange
multiplier \cite{cjp1, cjp2, cjp4, EPL}. Thus, the first law of
thermodynamics is valid for black strings, even after thermal
fluctuations are considered.

\section{ Conclusions}
In this paper, we will analyze a rotating black string in the
massive theory of  gravity. In this massive theory of gravity, the
gravitons have a mass, and these massive gravitons modify the
black string  solution. We will study thermal stability, critical
behavior and phase transition for such a black string by applying
different methods. We will also analyze the Van der Waals like
behavior for this solution. Finally, we will analyze the effects
of thermal fluctuations on the thermodynamics of such a solution.
It is known that in  the presence of electric charge, heat
capacity analysis is not sufficient to analyze the stability of
such a system. In fact, for such a system, the stability can be
studied using the entire Hessian matrix of the Helmholtz free
energy \cite{Hess1, Hess2, Hess3}.   Such a Hessian matrix is
obtained using   second derivatives of Helmholtz free energy with
respect to temperature and electric potential. Here we have used
such a Hessian matrix for analyzing the stability of black
strings.

It is interesting to note that there are other black brane
solutions, motivated by string theory \cite{1b, 2b, 4b, 5b}. It
would be interesting to analyze such solutions in massive gravity.
It would also be interesting to analyze the thermodynamics, and
Van der Waals behavior for such black branes solutions using
extended phase space. The stability of such black brane solutions
can again be analyzed using the Hessian matrix. It is expected
that here again the stability will depend on the mass of the
graviton. The quantum  fluctuations for such solutions can be
analyzed using thermal fluctuations to the thermodynamics of such
black holes \cite{cjp0, CJP}. It would also be interesting to
analyze the effect of such thermal fluctuations on the stability
of such black brane solutions.

\end{document}